\documentclass[final,3p,times]{elsarticle}

\usepackage{framed,multirow}

\usepackage{amssymb}
\usepackage{latexsym}

\usepackage[ruled,vlined]{algorithm2e}
\usepackage{physics}
\usepackage{bm}
\usepackage{subcaption}
\usepackage{bbold}
\usepackage{array}
\usepackage{mathtools}
\usepackage{epstopdf}
\usepackage{lineno}

\DeclareUnicodeCharacter{2212}{−}
\usepackage{url}
\usepackage{pgfplots}
\usepgfplotslibrary{groupplots,dateplot}
\pgfplotsset{compat=newest}
\usepackage{tikz}
\usetikzlibrary{patterns,shapes.arrows,spy,external,math}
\usepackage{algpseudocode}
\usepackage{verbatim}

\definecolor{crimson2143940}{RGB}{214,39,40}
\definecolor{darkgray176}{RGB}{176,176,176}
\definecolor{darkorange25512714}{RGB}{255,127,14}
\definecolor{forestgreen4416044}{RGB}{44,160,44}
\definecolor{lightgray204}{RGB}{204,204,204}
\definecolor{steelblue31119180}{RGB}{31,119,180}
\definecolor{darkblue}{RGB}{52,101,153}

\usepackage{anyfontsize}

\usepackage[normalem]{ulem}
\definecolor{readable_green}{RGB}{40, 189, 16}

\journal{Journal of Computational Physics}

\begin{document}

\begin{frontmatter}

\title{A fourth order sharp immersed method for the incompressible Navier-Stokes equations with stationary and moving boundaries and interfaces}

\author[MIT]{Xinjie Ji}
\author[MIT]{Changxiao Nigel Shen}
\author[MIT]{Wim M. van Rees\corref{cor1}}
\cortext[cor1]{Corresponding author. E-mail address: wvanrees@mit.edu}

\affiliation[MIT]{
    organization={Department of Mechanical  Engineering, Massachusetts Institute of Technology},%
    addressline={77 Masschusetts Ave.}, 
    city={Cambridge},
    postcode={02139}, 
    state={MA},
    country={United States}
}

\begin{abstract}

\noindent We propose a fourth order  Navier-Stokes solver based on the immersed interface method (IIM), for flow problems with stationary and one-way coupled moving boundaries and interfaces. Our algorithm employs a Runge-Kutta-based projection method that maintains high-order temporal accuracy in both velocity and pressure for steady and unsteady velocity boundary conditions. Fourth order spatial accuracy is achieved through a novel fifth order IIM discretization scheme for the advection term, as well as existing high-order interface-corrected finite difference schemes for the other differential operators. Using a set of manufactured flow problems with stationary and moving boundaries, we demonstrate fourth order convergence of velocity and pressure in the infinity norm, both inside the domain and on the immersed boundaries. The solver's performance is further validated through a range of practical flow simulations, highlighting its efficiency over a second order scheme. Finally, we showcase the ability of our immersed discretization scheme to handle interface-coupled multiphysics problems by solving a conjugate heat transfer problem with multiple immersed solids. Overall, the proposed approach robustly combines the efficiency of high order discretization schemes with the flexibility of immersed discretizations for flow problems with complex, moving boundaries and interfaces.

\end{abstract}

\end{frontmatter}

\section{Introduction}

Immersed methods are commonly used for solving fluid-structure interaction (FSI) problems due to their flexibility in enforcing boundary conditions on moving, deforming and complex geometries without the need to create and maintain body-fitted meshes \cite{griffith2020immersed}. Beyond FSI, these methods have been successfully applied to a range of multiphysics problems such as conjugate heat transfer \cite{koponen_direct_2025}, aeroacoustics \cite{jeong_immersed_2021, sabatini_arbitrary-order_2023}, and multiphase flows \cite{patel_diffuse_2018}. Among various formulations, sharp immersed methods are appealing as they resolve the immersed geometry without smearing or smoothing, and can resolve on- and near-surface quantities locally with high fidelity. %
Though the original immersed boundary method was first order accurate \cite{Peskin:1977}, many second-order incompressible Navier-Stokes solvers have since been developed. These typically use sharp boundary discretizations, such as the sharp interface immersed boundary method~\cite{Mittal2008}, the hybrid Cartesian/immersed boundary method~\cite{gilmanov_hcib_2005}, and the immersed interface method~\cite{Xu2006}.
However, high-order (greater than second) solvers incorporating immersed methods are significantly rarer, despite the potential benefits of high-order schemes in terms of efficiency and computational complexity~\cite{wang_high-order_2013}. Examples of high-order immersed schemes are \cite{Linnick2005,hosseinverdi_very_2017}, who developed fourth-order incompressible Navier-Stokes solvers in two dimensions (2D) using the vorticity-velocity formulation combined with an immersed interface method. While computationally efficient in 2D, such vorticity-velocity formulations are significantly more complex to extend to three-dimensional (3D) applications due to the increased number of Poisson solvers required for solving the velocity field, and the need to maintain a divergence-free vorticity field. In the velocity-pressure formulation, \citet{zhu_high-order_2016} used a compact finite difference immersed-boundary method to achieve up to fourth-order accuracy for velocity in 2D domains. Both these high-order approaches were only applied to domains with stationary immersed boundaries. To the best of our knowledge, no existing sharp immersed finite difference methods achieve high order accuracy for both velocity and pressure, or achieve high order in the presence of moving boundaries or immersed physical interfaces. This work aims to address this challenge by building upon a previously proposed collocated high-order Runge-Kutta based immersed interface method framework~\cite{gabbard2023high}. Our goal is to develop this framework into a high-order Navier-Stokes velocity-pressure algorithm for flows with stationary and one-way coupled moving immersed boundaries and interfaces.

To discuss relevant literature on high-order Navier-Stokes discretizations, it is useful to consider methods without immersed boundaries too. The reason is that our immersed interface scheme, like many other sharp immersed methods, constructs ghost point values by evaluating polynomials that directly incorporate the mathematical boundary conditions. This means the near-boundary discretizations are effectively linear combinations of domain values, or one-sided finite difference schemes, which is conceptually similar to how grid-aligned domain boundaries are discretized in standard finite-difference methods. The immersed scheme simply provides an algorithm to construct these schemes on-the-fly for arbitrarily shaped boundaries that do not necessarily align with the grid. As a result, any PDE formulation that provides conditions on domain-aligned boundaries can directly be applied to immersed boundaries as well, and vice versa. Therefore, we do not restrict our discussion below to immersed methods.

Navier-Stokes discretizations with or without immersed boundaries typically rely on fractional step techniques, originating from \cite{harlow_1965, chorin_numerical_1968, temam_sur_1969}. These algorithms are most often used with a staggered grid layout, so that the solutions satisfy discrete mass conservation without requiring explicit pressure boundary conditions. Such approaches are widely used in the immersed boundary community, e.g.\ \cite{Xu2006, Taira2007,Weymouth2025}, though projection-based collocated-grid methods have also been proposed \cite{Min2006,Mittal2008, blomquist_stable_nodal_2024}. 
The original projection method yields a pseudo-pressure that is only first-order accurate in time relative to the true pressure \cite{chorin_convergence_1969, perot_analysis_1993}. Strategies for improving pressure accuracy focus on designing projection methods that achieve higher-order pseudo-pressure accuracy, extensively discussed in the review of \citet{guermond_overview_2006}. Such approaches are often tied to specific temporal and spatial discretization schemes, and are challenging to directly translate to the context of collocated grids with explicit, high order time integration. 

In~\citet{zheng_rungekuttachebyshev_2006} the pressure accuracy of a Runge-Kutta based projection method was increased to second order by solving an additional pressure Poisson equation at each time. 
\citet{sanderse_accuracy_2012} formulated an approach that achieves second-order accurate pressure at the end of each time step by linearly combining pseudo-pressure results from the individual stages, avoiding the extra Poisson equation. Moreover, they analyzed Runge-Kutta-based fractional step schemes and observed that, for steady boundary conditions, the pseudo-pressure at the first stage of the next time step could provide a higher-order estimate of the current pressure. This insight was independently used to analyze pressure behavior near no-slip walls~\cite{vreman_projection_2014}. Recently, \citet{karam_high-order_2022} proposed a different approach for obtaining high-order pressure estimates, relying on linearly combining pseudo-pressures from different time levels. Such a recombination strategy offers an efficient way of achieving high-order pressure within projection methods. 

Lastly, we briefly mention the class of PPE (Pressure Poisson Equation) discretizations as an alternative to fractional step methods. PPE reformulations replace the incompressibility constraint with a Poisson equation, allowing velocity integration followed by pressure reconstruction~\cite{gresho_pressure_1987}. These methods can achieve high order accuracy for velocity and pressure \cite{johnston_accurate_2004} but typically require an additional boundary condition to ensure incompressibility \cite{petersson_stability_2001,rempfer_boundary_2006}. \citet{shirokoff_efficient_2011} combined PPE methods with an embedded boundary method for solving flow past static obstacles on a staggered grid. PPE-based approaches have also been explored on collocated grids and overlapping grids \cite{henshaw_fourth-order_1994, meng_fourth-order_2020}. A main drawback of PPE methods, especially on collocated grids, is that the incompressibility is not explicitly enforced; in practice, many methods rely on ad-hoc damping techniques to dynamically dissipate spurious divergence in the flow. 

In this work, we propose a variation of the Runge-Kutta projection scheme of \citet{sanderse_accuracy_2012} that explicitly formulates the pseudo-pressure boundary conditions so that the temporal order of accuracy of the pressure matches that of the velocity. By integrating this scheme with our sharp immersed method \cite{gabbard2023high,gabbard_2024}, we achieve fourth-order accuracy in solving the incompressible Navier-Stokes equations with static and moving embedded boundaries. We highlight the flexibility of the immersed method by solving external and internal flow problems, as well as conjugate heat transfer with immersed interfaces. 

The remainder of this paper is structured as follows: Section~\ref{sec: MethodsP1} reviews the immersed interface method (IIM) and the discretization schemes used in our solver, where a novel fifth-order advection scheme is proposed. Section~\ref{sec: MethodsP2} discusses the Runge-Kutta based projection algorithm for the Navier-Stokes equations, and its integration with our immersed discretization methods for stationary and moving boundaries. Section~\ref{sec: resultsP1} verifies the accuracy and stability of the resulting fourth order scheme through comparisons with exact solutions. To validate, we apply the algorithm to benchmark flow simulations and extend it to a conjugate heat transfer example in Section~\ref{sec: resultsP2}, comparing the proposed fourth order scheme with a second order one. Finally, we present our conclusions and discuss potential future extensions in Section~\ref{sec: conclusion}.

\section{High order IIM discretization of advection-diffusion problems} \label{sec: MethodsP1}
In this section, we first present our high order immersed interface method (IIM) for the discretization of advection-diffusion problems with immersed stationary and moving boundaries. This serves to provide the necessary background and introduce notation for the Navier-Stokes discretization discussed in section~\ref{sec: MethodsP2}. Specifically, subsections~\ref{subsec:FD_IIM} and \ref{subsec: IIM moving}, review our previous work on the IIM for stationary and moving boundaries, respectively \cite{gabbard2023high,gabbard_2024}. Subsequently, subsection~\ref{sec: stabilityerror} introduces and verifies a novel, fifth-order accurate discretization of the advection term.

\subsection{Immersed finite-difference discretization}
\label{subsec:FD_IIM}
The immersed method provides local corrections to standard finite difference schemes to incorporate boundary or interface conditions. Here we use dimension-split finite difference stencils on a uniform Cartesian grid with spacing $h$. Away from immersed geometries, the finite-difference discretizations are standard centered finite-difference schemes of second or fourth order for all first and second derivatives, except for advection terms; advection terms are discretized using standard upwind schemes of third or fifth order.

Conventional finite difference stencils require corrections whenever they intersect immersed boundaries or interfaces. The boundary corrections used in this work are based on the immersed interface method \cite{Leveque1994, Li2006, Wiegmann2000}, substantially simplified using polynomial extrapolation \cite{Gabbard2021}. Following a convention from the immersed interface literature, we refer to the intersection between a grid-line and the surface $\Gamma$ as a \textit{control point}, denoted $\mathbf{x}_c$. Any evaluation point for a finite-difference stencil intersecting the surface is referred to as an \textit{affected point}, since the discretization is affected by the presence of the surface. To discuss their treatment, we first consider immersed boundaries, where the PDE is posed in domain $\Omega^+$; subsequently we extend to immersed interfaces where PDEs are posed on both $\Omega^+$ and $\Omega^-$, and coupled across the immersed interface $\Gamma$. 

\subsubsection{Immersed boundaries}
Each control point $\mathbf{x}_c$ on an immersed domain boundary is associated with a set of interpolation points 
$\mathcal{X}_c^+ \subset \Omega^+$, 
and with a multivariate polynomial $p_c(\mathbf{x})$ of degree $k$ that approximately interpolates the domain values $\qty{f(\mathbf{x}_{\alpha}) \mid \mathbf{x}_{\alpha} \in \mathcal{X}_c^+}$ in a least squares sense. Any 1D finite difference stencil that intersects the boundary at $\mathbf{x}_c$ is applied to the extended function
\begin{equation}\label{eq:extended_function}
    f_c(\mathbf{x}) = \begin{cases} f(\mathbf{x}), & \mathbf{x} \in \Omega^+\\ p_c(\mathbf{x}), & \mathbf{x} \notin \Omega^+
\end{cases}.
\end{equation}

\begin{figure}[tb!]
    \centering
    \begin{subfigure}{0.46\textwidth}
        \centering
        \resizebox{0.7\textwidth}{!}{
            \includegraphics[width=\textwidth]{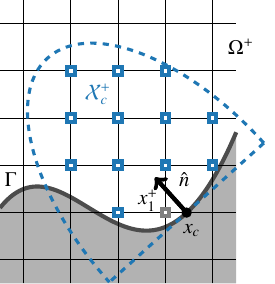}
        }
        \caption{Multidimensional interpolant for immersed boundaries.}
        \label{fig:iim_stencil_boundary}
    \end{subfigure}
    \hspace{0.05\textwidth}
    \begin{subfigure}{0.46\textwidth}
        \centering
        \resizebox{0.7\textwidth}{!}{
            \includegraphics[width=\textwidth]{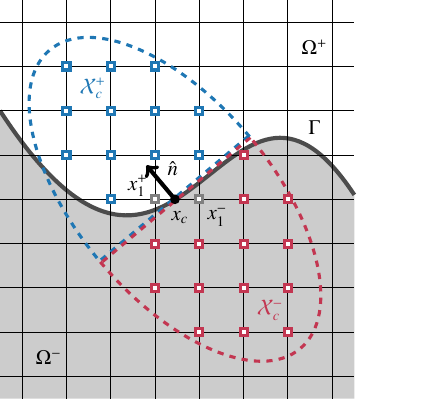}
        }
        \caption{Multidimensional interpolants for immersed interfaces.}
    \label{fig:iim_stencil_interface}
    \end{subfigure}
    \caption{Each crossing between a grid line and the boundary ($x_c$)  is used to construct ghosts points for the affected grid points ($x_1^+$,$x_1^-$) using a multidimensional interpolant constructed from a half-elliptical region of grid points. For immersed boundaries (a), the interpolant is constructed using imposed boundary conditions (Dirichlet, Neumann). For immersed interfaces (b), polynomials from both sides are constructed using imposed jump conditions.}
    \label{fig:iim_stencil}
\end{figure}

For control points with prescribed boundary conditions, the set of interpolation points in the least squares domain  $\mathcal{X}_c^+$ includes the control point, excludes the closest grid point, and includes all other grid points that are (1) part of the domain $\Omega^+$, and (2) fall within a half-elliptical region centered on the boundary whose semi-major axis is aligned with the local normal vector to the surface (Figure~\ref{fig:iim_stencil_boundary}). When a boundary condition is not prescribed, we modify $\mathcal{X}_c^+$ to omit the control point and instead include the closest grid point. The major and minor axes of the half-elliptical region are chosen so that we can guarantee the existence of $p_c(\mathbf{x})$ on a grid with spacing $h$ as long as the immersed surface satisfies
\begin{equation}
\label{eq:curvature_condition}
|\kappa_{max} h| < 1/4,
\end{equation}
where $\kappa_{max}$ is the maximum scalar curvature of the surface \cite{gabbard_2024}. Roughly, this leads to a semi-major axis of $\sim k$ grid points, and semi-minor axis of $\sim k/2$ grid points --- precise values are provided in \cite{gabbard2023high}.

Equation~\eqref{eq:extended_function} implies that $p_c(\mathbf{x})$ needs to be evaluated at grid points $\mathbf{x}_g$ that fall outside the domain. While it is possible to explicitly form the interpolant $p_c(\mathbf{x})$ and then evaluate it, in practice it is more convenient to write the desired quantities $p_c(\mathbf{x}_g)$ directly as linear combinations of the values of $f(\mathbf{x}_\alpha)$ at the interpolation points. Introducing stencil coefficients $\{s_c^g\} \cup \{s^g_\alpha\}_{\alpha = 1}^n$ and considering the case where a boundary condition is imposed, this yields
\begin{equation}
    p_c(\mathbf{x}_g) = s^g_c f(\mathbf{x}_c) + \sum_{\alpha = 1}^n s^g_\alpha f(\mathbf{x}_\alpha)
\end{equation}
at each point $\mathbf{x}_g$ requiring a ghost value \cite{gabbard_2024}. For control points with Dirichlet boundary conditions, $f(\mathbf{x}_c)$ is given by the boundary condition. For points with Neumann boundary conditions $f(\mathbf{x}_c)$ is not directly available, but it can be approximated based on the boundary condition and nearby solution values. To clarify, let $\{s_c\} \cup \{s_i\}_{i = 1}^n$ be a set of stencil coefficients that approximate the normal derivative of $p_c$ at $\mathbf{x}_c$, so that
\begin{equation}\label{eq:normal_grad}
    \partial_n f(\mathbf{x}_c) = s_c f(\mathbf{x}_c) + \sum_{i = 1}^{n} s_i f(\mathbf{x}_i) + \order{\Delta x^{k - 1}}.
\end{equation}
When a Neumann condition $\partial_n f = q$ is prescribed at a control point, Equation~\eqref{eq:normal_grad} can be inverted to give
\begin{equation}\label{eq:neumann}
    f(\mathbf{x}_c) = \frac{1}{s_c} \qty(q(\mathbf{x}_c) - \sum_{i = 1}^{n} s_i f(\mathbf{x}_i)) + \order{\Delta x^{k}}.
\end{equation}
This requires one additional set of stencil coefficients which evaluate the normal derivative $\partial_n p(\mathbf{x}_c)$ on the boundary. Lastly, in case no boundary conditions are specified (e.g.\ for outflow boundaries in advection problems), the polynomial is constructed without boundary value $f(\mathbf{x}_c)$, leaving the PDE unconstrained at the immersed geometry. We treat domain boundaries conceptually similar to IIM boundaries, except we use a 1D polynomial extrapolation that never omits the closest points.

The least-squares interpolant $p_c(\mathbf{x})$ is constructed using uniform weights for all finite-difference schemes except for the advection term. Previous work found that choosing weights which rapidly decay away from the wall enhances stability of high order advection discretizations in two and three dimensions \cite{devendran2017fourth, gabbard_2024}. These same weights are adopted here for constructing ghost points of advection terms near inflow boundaries (i.e.\ wherever $\mathbf{u}(\vb{x}_c) \cdot \mathbf{n}(\vb{x}_c) \geq 0$ with $\mathbf{n}(\vb{x}_c)$ the unit normal pointing to the flow domain); for outflow boundaries, we rely on polynomials constructed without boundary conditions. 

\subsubsection{Immersed interfaces}
The immersed boundary formulation presented above is easily extended to interface-coupled multiphysics problems, such as conjugate heat transfer. In such problems, different PDEs hold on either side of the interface, which are typically coupled by interface-jump conditions prescribed on the solution and its flux. Denoting the interface between the subdomains $\Omega^+$ and $\Omega^-$ as $\Gamma$, these conditions can be written as
\begin{equation}\label{eq:poisson_jumps}
\begin{aligned}
    [f](s) &= j_0(s)\; \text{on} \; \Gamma, \\
    [\kappa \partial_n f](s) &= j_1(s) \; \text{on} \; \Gamma,
\end{aligned}
\end{equation}
where $s$ is the boundary coordinate and $\kappa$ is the problem-dependent and typically discontinuous transport coefficient. To discretize these jump boundary conditions, the boundary values $f^-(\mathbf{x}_c)$ and $f^+(\mathbf{x}_c)$ from either side of the interface are computed with the aid of two sets of stencils coefficients, each associated with their own half-elliptical regions as shown in Figure~\ref{fig:iim_stencil_interface}. The first set $\{s^+_c, s^+_i\}$ maps the boundary value $f^+(\mathbf{x}_c)$ and solution values from $\Omega^+$ to the normal derivative $\partial_n f^+(\mathbf{x}_c)$, while the second set $\{s_c^-, \, s_i^-\}$ is designed analogously to map solution values from $\Omega^-$ to the normal derivative $\partial_n f^-(\mathbf{x}_c)$. The boundary values $f^\pm(\mathbf{x}_c)$ can then be determined from the discretized jump conditions $f^+(\mathbf{x}_c) - f^-(\mathbf{x}_c) = j_0(\mathbf{x}_c)$ and
\begin{equation}
\label{eq:boundary_system_scalar}
        \kappa^+\qty(s^+_c f^+(\mathbf{x}_c) + \sum_{i=1}^{n^+} s^+_i f(\mathbf{x}_i^+)) - \kappa^-\qty(s^-_c f^-(\mathbf{x}_c) + \sum_{i=1}^{n^-} s^-_i f(\mathbf{x}_i^-)) = j_1(\mathbf{x}_c).
\end{equation}
Once determined, the boundary values $f^\pm(\mathbf{x}_c)$ can be used in stencil operations on either side of the interface, similar to the procedure outlined above for domain boundary conditions. For more details we refer to \cite{gabbard2023high}.

\subsection{IIM moving boundary treatment}
\label{subsec: IIM moving}
With moving boundaries, grid points may enter or exit the PDE domain during the time integration. Points exiting the domain can be ignored in the subsequent spatial discretizations, but points newly entering the domain need to be included in the solution process. However, since their time history is unavailable, it is challenging to achieve high order time integration with moving immersed boundaries. To address this, we briefly recall the high-order method proposed in \cite{ji_sharp_2023} for low-storage Runge-Kutta (LSRK) methods \cite{williamson_low-storage_1980}. We present the method in the context of a general PDE of the form $\partial f / \partial t = g(f, t)$, discretized using an $s$-stage LSRK integrator with a time step size $\Delta t$~\cite{williamson_low-storage_1980}. The extension to the Navier-Stokes equations is discussed in subsection~\ref{subsec: NS moving bc}. We denote the boundary as $\Gamma(t)$. Our immersed method typically relies on a level-set field $\psi(\vb{x},t)$, from which instantaneous control point locations and normal vectors can be computed at any instance in time \cite{gabbard2023high, gabbard_2024}; other geometric representation are also possible. In this work, we only consider domains with prescribed motion where $\Gamma(t)$ is given explicitly and hence does not need to be numerically integrated in time.

We start from a general LSRK time integration of $f_t = g(f,t)$:
\begin{align}
    &v^{(i)} = \hat{a}_i v^{(i-1)} + \Delta t g \left( f^{(i-1)}, t^{(i-1)} \right), \\
    &f^{(i)} = f^{(i-1)} + \hat{b}_i v^{(i-1)},\\
    &t^{(i)} = t^{(i-1)} + c_i \Delta t,
\end{align}
where $1 \leq i \leq s$ denotes the RK stage index. The coefficients $\hat{a}_i$, $\hat{b}_i$, and $c_i$ are predefined constants specific to the LSRK scheme. The variables $f^{(i)}$, $v^{(i)}$ and $t^{(i)}$ represent the solution, the intermediate variable and time at stage $i$, respectively, while $f^{(0)}$, $v^{(0)}$ and $t^{(0)}$ denote their values from the previous time step. 
To adapt the IIM discretizations with moving boundaries into the general PDE formulation, extrapolations are performed at each LSRK stage to provide valid values for grid points newly entering the fluid domain. Specifically, given the boundary $\Gamma^{(i-1)} = \Gamma(t^{(i-1)})$, we extrapolate both $f^{(i-1)}$ and $g(f^{(i-1)}, t^{(i-1)})$ one layer of grid points beyond the boundary, i.e.\ inside the body. Analogously to \cite{ji_sharp_2023,gabbard_2024} we define these extrapolation operations through $E_{D}^{(i-1)}[\cdot]$ and $E_{F}^{(i-1)}[\cdot]$, representing polynomial extrapolation with Dirichlet and free boundary conditions, respectively. These operators evaluate the same least squares multivariate polynomials as described in section~\ref{subsec:FD_IIM} above. We further define a zeroing operator $Z^{(i)}[\cdot]$, which sets the solution to zero within the body region defined by $\Gamma^{(i)} = \Gamma(t^{(i)})$. The modified LSRK scheme incorporating the IIM moving boundary treatment then becomes:

\begin{align}
    &v^{(i)} = \hat{a}_i v^{(i-1)} + \Delta t E_{F}^{(i-1)}\left[g \left( f^{(i-1)}, t^{(i-1)} \right)\right], \\
    &f^{(i)} = Z^{(i)}\left[E_{D}^{(i-1)}\left[f^{(i-1)}\right] + \hat{b}_i v^{(i-1)}\right],\\
    &t^{(i)} = t^{(i-1)} + c_i \Delta t.
\end{align}

As shown in~\cite{ji_sharp_2023,gabbard_2024}, when $g$ arises from a IIM-based spatial discretization this method achieves convergence with an overall error given by
\begin{equation}
    \| f - f_e\| = \mathcal{O}{(\Delta t^\gamma)} + \mathcal{O}{(h^{\eta})} + \mathcal{O}(\Delta t h^{\Tilde{\eta}}),
\end{equation}
where $\gamma$ is the temporal order of the RK integrator, $\eta$ is the spatial accuracy of the discretization, and $\Tilde{\eta}$ depends on both the spatial discretization and the order of the IIM extrapolation. Although a mixed space-time error term $\mathcal{O}(\Delta th^{\Tilde{\eta}})$ arises, it does not affect the overall accuracy due to the body-CFL constraint. In particular, the body-CFL constraint requires that $\Delta t \sim h$, so the mixed error term is still a high order error term. For more discussion about this mixed error term, and the extension of this approach to moving interface problems, we refer to~\cite{gabbard_2024}. 

\subsection{Novel fifth-order IIM advection scheme}\label{sec: stabilityerror}
In previous work \cite{gabbard2023high, gabbard_2024} we combined fourth order polynomials with an upwind third order finite difference scheme to achieve up to third-order accuracy for discretizations of the transport equation with immersed boundaries. It was also shown that extending this strategy to sixth order polynomials with an upwind fifth order finite difference scheme leads to an unconditionally unstable discretization. In fact, we are not aware of higher than third order inflow treatments that remain stable using immersed finite difference schemes, besides those resorting to inverse Lax-Wendroff boundary treatments~\cite{sirui_2021, jianfang_2021}; unfortunately inverse Lax-Wendroff-based strategies are challenging to extend to complex PDEs like the incompressible Navier-Stokes equations.

In this section we address this gap by presenting a novel, conditionally stable fifth-order immersed advection scheme, which is incorporated in the Navier-Stokes discretization presented below. 

\subsubsection{Analysis} 
The new fifth-order scheme relies on a sixth-order extrapolation technique to fill in ghost points across an immersed inflow boundary, using the boundary condition itself in the interpolant construction. As mentioned above, one could try to apply the standard fifth order upwind finite difference scheme to this extended field
\begin{equation}
    \left(\frac{\partial f}{\partial x}\right)_k =
    \begin{dcases}
     \frac{-2f_{k-3}+15f_{k-2}-60f_{k-1}+20f_{k}+30f_{k+1}-3f_{k+2}}{60h} + \mathcal{O}(h^5) & u_k \geq 0,\\
     \frac{3f_{k-2}-30f_{k-1}-20f_{k}+60f_{k+1}-15f_{k+2}+2f_{k+3}}{60h} + \mathcal{O}(h^5) & u_k < 0.
    \end{dcases}
    \label{eq: advection discretization 5free}
\end{equation} 

\begin{figure}
    \begin{subfigure}{0.34\linewidth}
        \centering
        \resizebox{\textwidth}{!}{\includegraphics[width=1.2\textwidth]{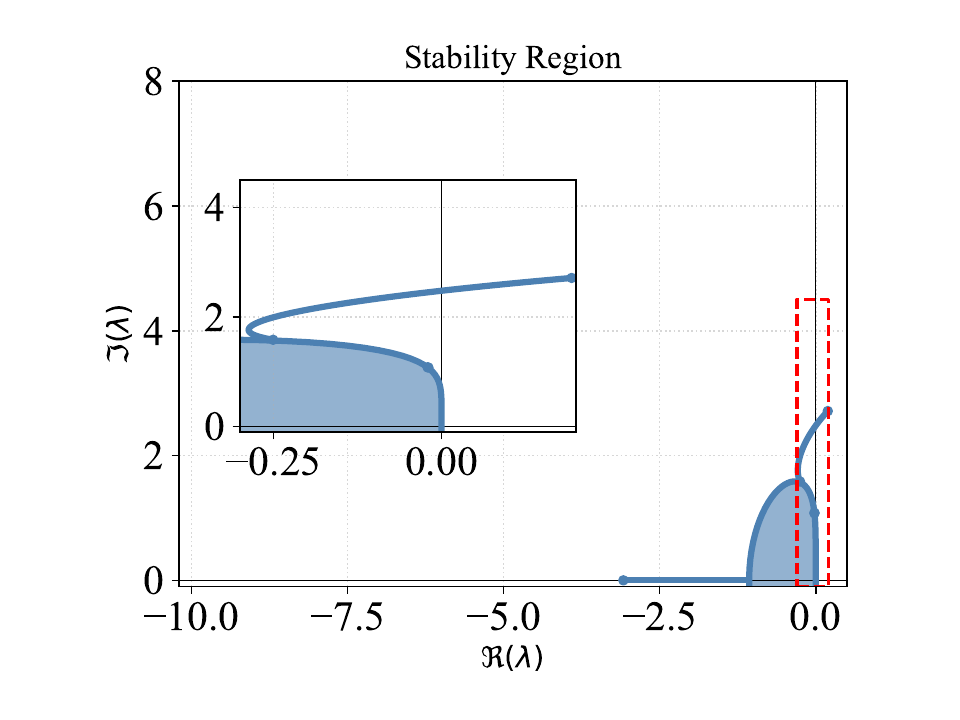}}
        \caption{Stability region of the normal fifth-order upwind stencil.}
    \end{subfigure}
    \hspace{0.0\textwidth}
    \begin{subfigure}{0.34\linewidth}
        \centering
        \resizebox{\textwidth}{!}{\includegraphics[width=1.2\textwidth]{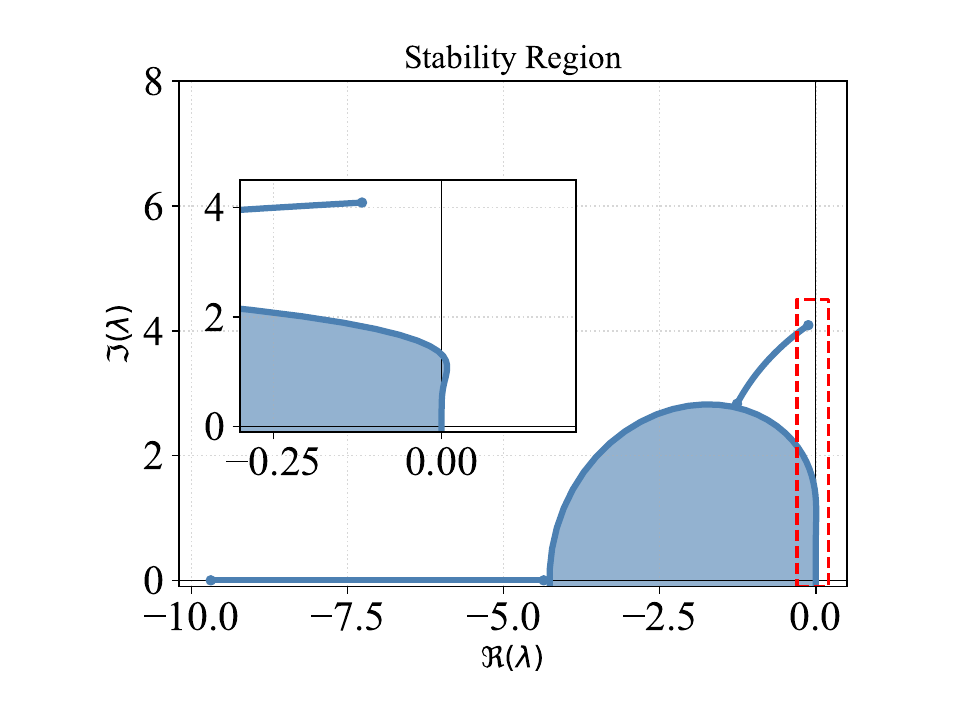}}
        \caption{Stability region of the extreme fifth-order upwind stencil.}
    \end{subfigure}
    \hspace{0.0\textwidth}
    \begin{subfigure}{0.34\linewidth}
        \centering
        \resizebox{\textwidth}{!}{\includegraphics[width=1.2\textwidth]{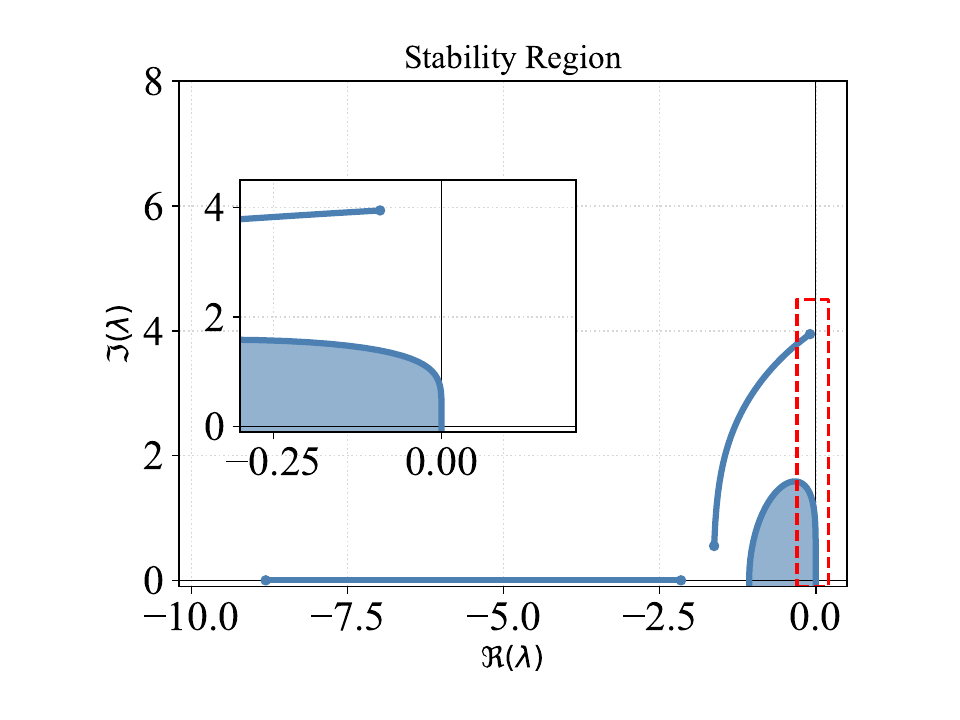}}
        \caption{Stability region of the mixed fifth-order upwind stencil.}
    \end{subfigure}
    \caption[]{Stability regions of the normal fifth-order upwind stencil in Equation \ref{eq: advection discretization 5free}, the fifth-order extreme upwind stencil in Equation \ref{eq: advection discretization 5}, and the mixed fifth order scheme. The filled region indicates the eigenvalues of the free-space discretization, and solid lines represent the eigenvalues associated with the IIM boundary treatment, sweeping over all possible intersection locations. Results are obtained via a GKS stability analysis. } 
    \label{fig: stability1}
\end{figure}

However, this approach was shown to be unconditionally unstable using a GKS stability analysis \cite{gustafsson1972stability} of a 1D transport problem in a semi-infinite domain \cite{gabbard2023high}. The associated stability region is reproduced here in Figure~\ref{fig: stability1}a, with the free-space discretization shown as the shaded region and the IIM-related eigenvalues shown as the solid lines. These solid lines represent solutions obtained by systematically varying the location of the interface between two grid points, so that all possible interface locations are covered. The plot shows that eigenvalues associated with specific intersection locations cross into the positive real plane, leading to unconditional instability. 

To mitigate this, we consider the following fifth-order `extreme' upwind scheme:
\begin{equation}
   \left(\frac{\partial f}{\partial x}\right)_k =
   \begin{dcases}
    \frac{3f_{k-4} - 20f_{k-3} + 60f_{k-2} - 120f_{k-1} + 65f_k + 12f_{k+1}}{60h} + \mathcal{O}(h^5) & u_k \geq 0,\\
    \frac{-12f_{k-1} - 65f_{k} + 120f_{k+1} - 60f_{k+2} + 20f_{k+3} - 3f_{k+4}}{60h} + \mathcal{O}(h^5) & u_k < 0.
   \end{dcases}
   \label{eq: advection discretization 5}
\end{equation} 
The 1D stability region of this scheme, combined with a sixth order IIM polynomial, is shown in Figure~\ref{fig: stability1}b. The plot shows that the extreme eigenvalues associated with the immersed inflow boundary (solid lines) now do not acquire positive real parts, but the stability region of the free-space stencil (shaded region) does cross into the positive real plane.

We therefore propose a mixed scheme in which the extreme upwind stencil is used only near domain boundaries (within a distance of $3h$), while the normal upwind stencil is applied in the interior. This approach guarantees formal stability through the 1D GKS stability analysis, as illustrated in Figure \ref{fig: stability1}c. In particular, the free-space stability region of our mixed scheme matches that of the normal fifth-order scheme, since both use identical free-space finite difference stencils.  Near boundaries, we switch to the 'extreme' upwind stencil with ghost points evaluated from the sixth order polynomial interpolant, which ensures that all eigenvalues from the IIM discretization have negative real parts. 

A drawback of this approach is that the local truncation error (LTE) of the extreme upwind stencil is twice larger than for the normal upwind stencil. However, we consider this trade-off justified by the improved stability of the fifth-order scheme for advection problems, and the fact that this error is only incurred near the boundaries.

\subsubsection{Numerical convergence}
\label{subsec:fifthorder_convergence}
To assess the convergence properties of the proposed mixed fifth-order advection scheme we apply it to the 2D linear advection equation
\begin{equation}
    \frac{\partial f}{\partial t} + \vb{u} \cdot \nabla f = 0.
\end{equation}
The computational domain is a uniform Cartesian grid over the unit square with $N^2$ grid points. The immersed boundary is an embedded star-shaped geometry whose boundary is parametrically defined as
\begin{equation}
    r(\theta) = r_b + r_d \cos (N_s \theta), \label{eq: star-like levelset}
\end{equation}
where $(r, \theta)$ are polar coordinates relative to the body center $\vb{x}_b = [0.51, 0.52]$, with $r_b$ as the base radius, $r_d$ as the deviation radius, and $N_s = 5$ determining the number of star corners. Two distinct test cases are simulated until $t = 2.0$ with LSRK4 time integration~\cite{kennedy_low-storage_2000}. A fixed CFL condition of \(\Delta t/h=0.1\) is maintained across all simulations.

\begin{figure}
    \begin{subfigure}[t]{0.25\linewidth}
        \centering
        \includegraphics[width=1\textwidth]{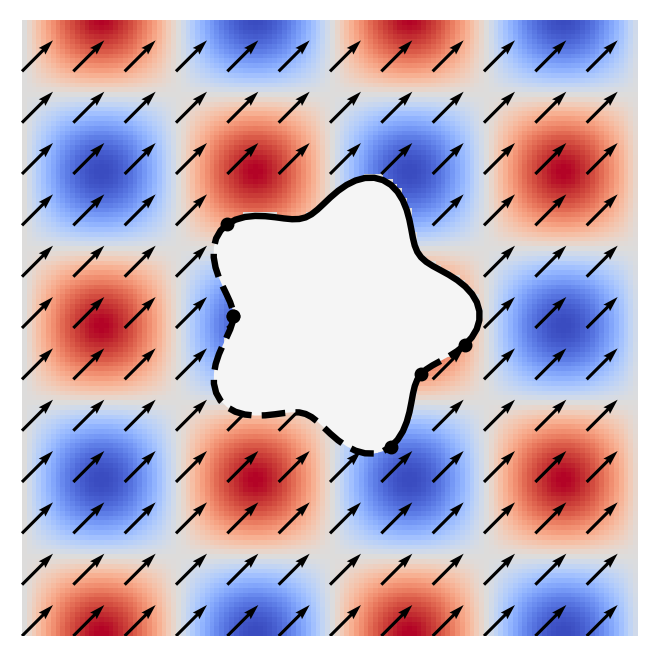}
        \caption{Initial setup.}
        \label{fig: adv test1 initial}
    \end{subfigure}
    \hspace{0.01\textwidth}
    \begin{subfigure}[t]{0.36\linewidth}
        \centering
        \resizebox{\textwidth}{!}{\includegraphics[width=1\textwidth]{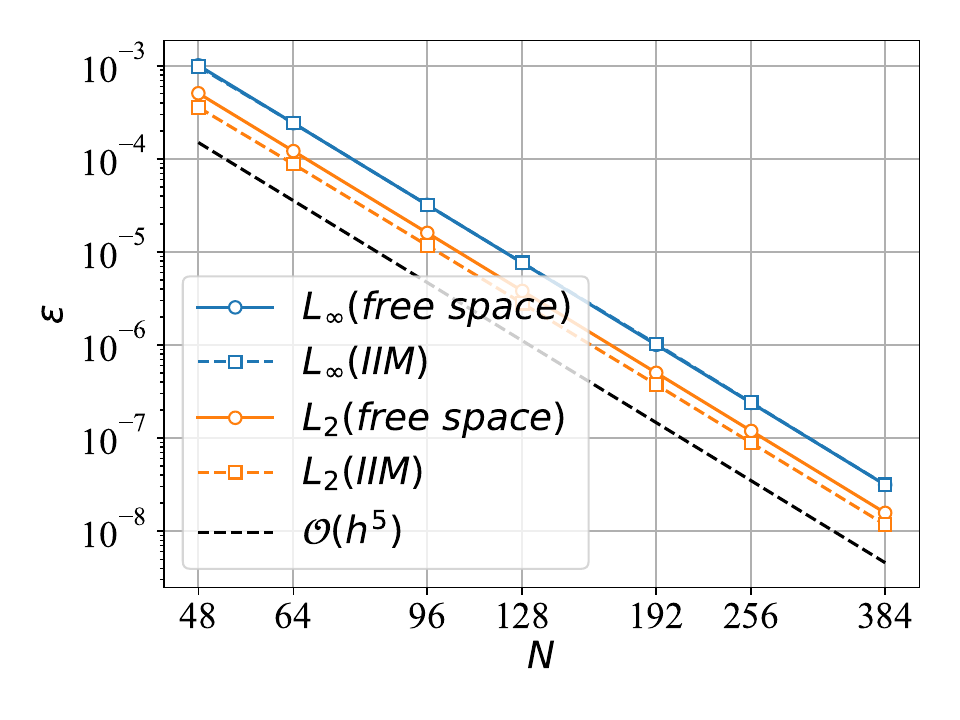}}
        \caption{\(L_2\) and \(L_\infty\) error vs. spatial resolution.}
    \end{subfigure}
    \hspace{0.01\textwidth}
    \begin{subfigure}[t]{0.3\linewidth}
        \centering
        \includegraphics[width=1\textwidth]{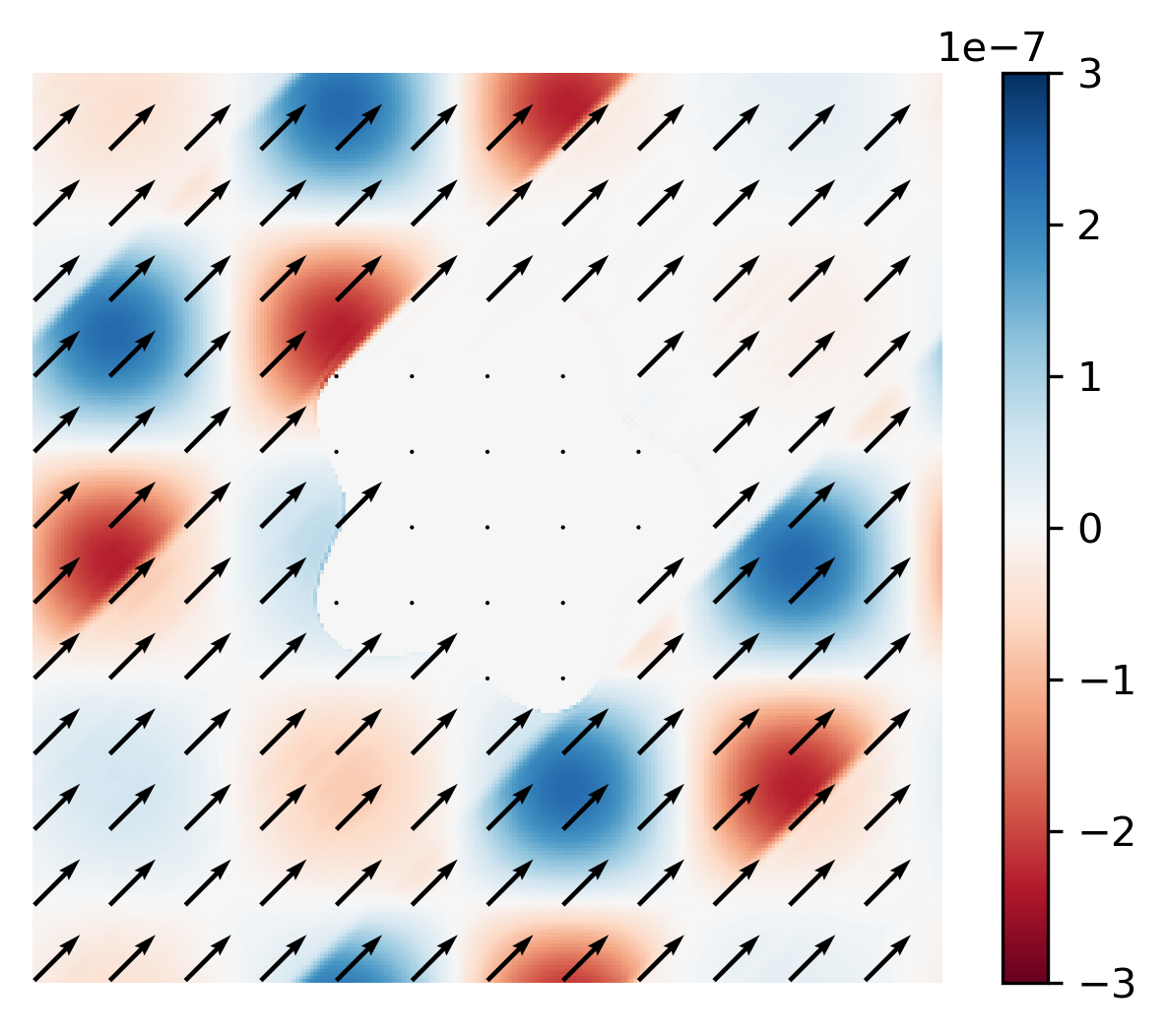}
        \caption{Final time error field for case with an embedded obstacle with \(N^2=256\).}
    \end{subfigure}
    \caption[]{Linear advection test case with a translational flow in a periodic domain with (IIM) and without (free space) an embedded star-shaped obstacle.}
    \label{fig: adv_conv1}
\end{figure}

\paragraph{Test Case 1: Uniform Flow in a Periodic Domain}  
In the first test, the velocity field is uniform, given by \(\vb{u} = [1.0, 1.0]\), and periodic boundary conditions are imposed on all sides of the domain. The star-shaped obstacle is embedded with $r_b = 0.2$ and $r_d = 0.033$, as shown in Figure~\ref{fig: adv test1 initial}. The exact solution is given by
\begin{equation}
    f(\vb{x},t) = \sin(4\pi (x-t))\cos(4\pi (y-t)),
\end{equation}
with $\vb{x} = (x,y)^T$. For this test case we consider the errors of the proposed mixed fifth-order upwind IIM scheme (with the immersed boundary) and the normal upwind stencil in Equation~\eqref{eq: advection discretization 5} (where the embedded star is removed to ensure stability). Figure~\ref{fig: adv_conv1}b and~\ref{fig: adv_conv1}c shows the \(L_2\) and \(L_\infty\) norm errors of the numerical solution \(f\) versus $N$. Both schemes show clear fifth-order convergence, verifying the accuracy of our schemes in both scenarios. Further, there is no noticeable difference between the $L_{\infty}$ error of the two cases (the lines overlap), indicating that for this test case the largest errors appear away from the boundary.

\begin{figure}
    \begin{subfigure}{0.25\linewidth}
        \centering
        \includegraphics[width=1\textwidth]{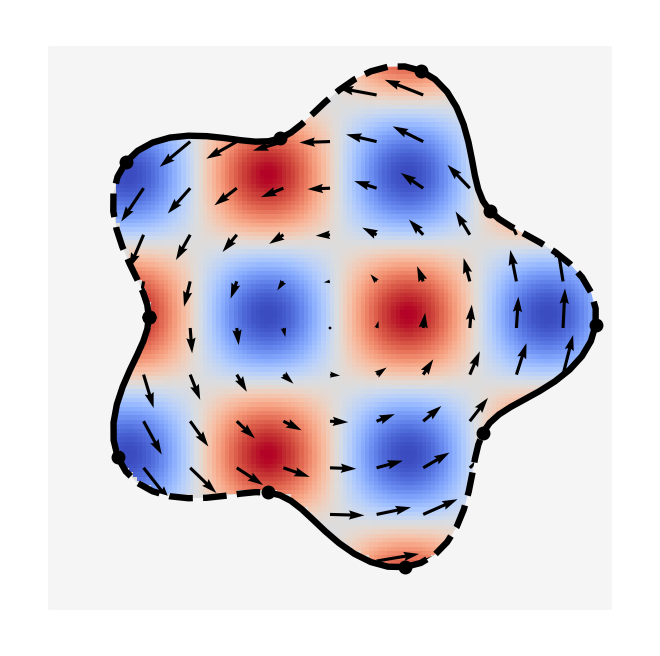}
        \caption{Initial setup.}
        \label{fig: adv test 2 illu}
    \end{subfigure}
    \hspace{0.01\textwidth}
    \begin{subfigure}{0.36\linewidth}
        \centering
        \resizebox{\textwidth}{!}{\includegraphics[width=1\textwidth]{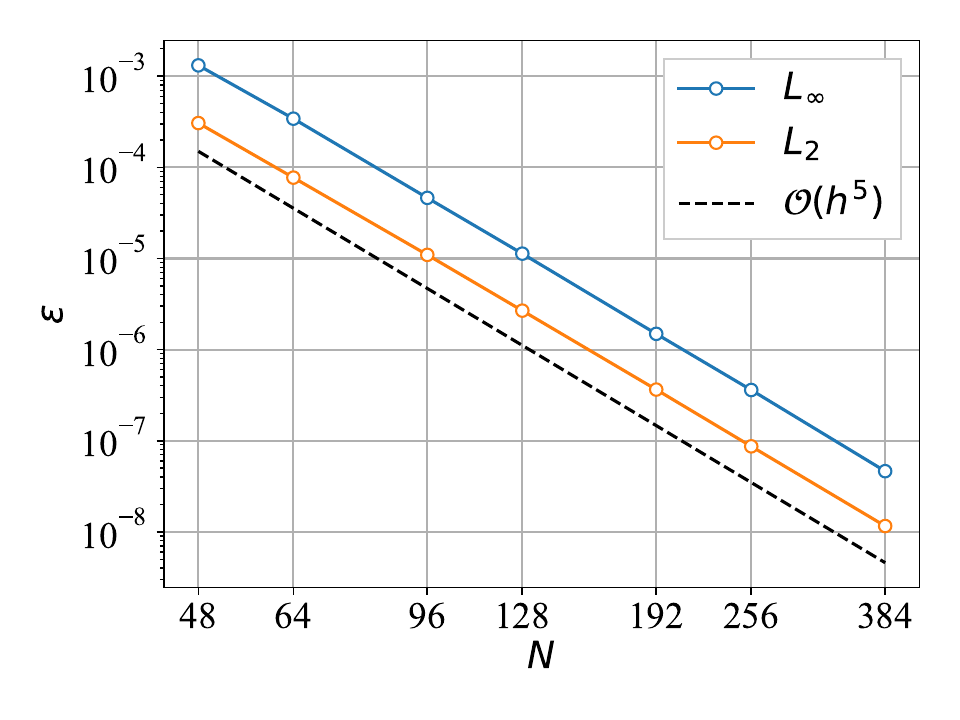}}
        \caption{\(L_2\) and \(L_\infty\) error vs. spatial resolution.}
    \end{subfigure}
    \hspace{0.01\textwidth}
    \begin{subfigure}{0.3\linewidth}
        \centering
        \includegraphics[width=1\textwidth]{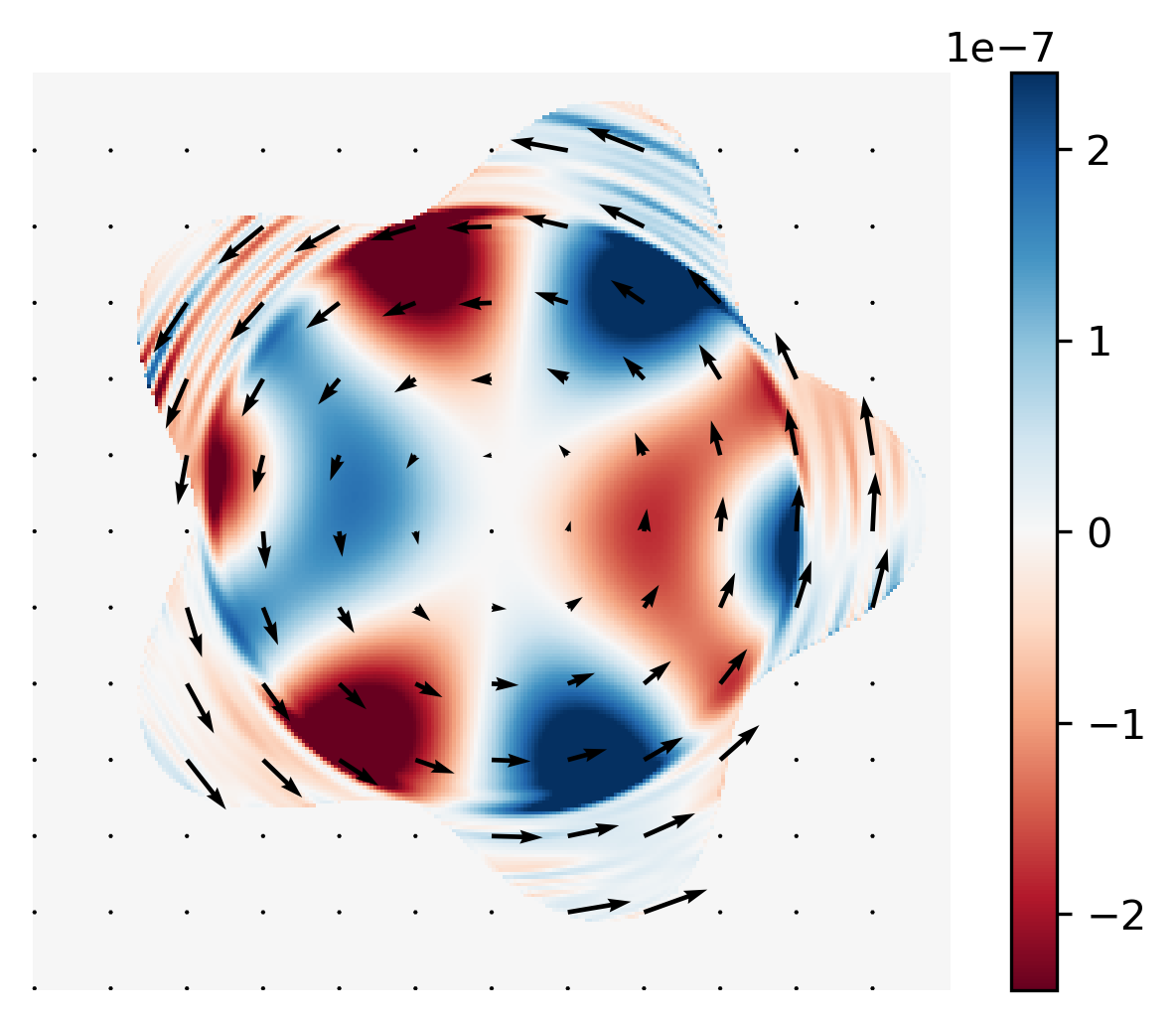}
        \caption{Final time error field for \(N=256^2\).}
    \end{subfigure}
    \caption[]{Linear advection test case with a rotational flow in a star-shaped domain.}
    \label{fig: adv_conv2}
\end{figure}

\paragraph{Test Case 2: Rotating Flow in a Star-Shaped Domain}  
For the second test, we simulate a rotational velocity field within a closed star-shaped domain, specified by $r_b = 0.4$ and $r_d = 0.066$ (Figure~\ref{fig: adv test 2 illu}). The advection velocity is now defined as
\begin{equation}
    \textbf{u} = [y_0-y, x-x_0],
\end{equation}
with the rotation center located at \([y_0, x_0] = [0.51, 0.52]\). The exact solution takes the form:
\begin{equation}
    f(\vb{x},t) = \sin\left(4\pi p_x(\vb{x},t)\right) \cos\left(4\pi p_y(\vb{x},t)\right)
\end{equation} where \begin{equation}
    \begin{bmatrix}
        p_x(\vb{x},t)\\
        p_y(\vb{x},t)
    \end{bmatrix} = \begin{bmatrix}
        \cos(\pi t) & \sin(\pi t)\\
        -\sin(\pi t) & \cos(\pi t)
    \end{bmatrix}\begin{bmatrix}
        x-x_0\\
        y-y_0
    \end{bmatrix}.
\end{equation}
Figure \ref{fig: adv_conv2}b shows the spatial convergence of the $L_2$ and $L_{\infty}$ norm errors for this case using the proposed mixed fifth-order scheme. The corresponding error field at the final simulation time is shown in Figure~\ref{fig: adv_conv2}c. The results demonstrate fifth-order convergence, consistent with the findings from the first test case. 

Overall, the novel fifth-order IIM advection scheme proposed in this section is shown to achieve comparable accuracy and stability to a standard free-space fifth-order upwind finite-difference stencil.

\section{IIM discretization of the incompressible Navier-Stokes equations}
\label{sec: MethodsP2}
In this section, we introduce a high-order projection method for solving the incompressible Navier-Stokes equations on collocated grids. After presenting the governing equations in section~\ref{subsec:NS-eqns}, we consider discretizations of static boundaries in \ref{subsec: NS discretizations}, and moving boundaries in \ref{subsec: NS moving bc}. Finally, subsection~\ref{subsec:implementation} discusses the implementation of the algorithm.  While the focus of this work is on the two-dimensional Navier-Stokes equations, the approach can be readily extended to three dimensions.

\subsection{Governing equations} 
\label{subsec:NS-eqns}
We consider the velocity-pressure formulation of the incompressible Navier-Stokes equations in a domain $\Omega$ with an arbitrary closed body boundary $\Gamma$. The flow field $\vb{u}(\vb{x}, t)$ and pressure field $p(\vb{x}, t)$ satisfy 

\begin{align}
    & \frac{\partial \vb{u}}{\partial t} + \vb{u} \cdot \nabla \vb{u} = -\nabla p + \nu \Delta \vb{u}, \label{eq: NS momentum}  &\quad\\
    & \nabla  \cdot \vb{u} = 0, &\vb{x} \in \Omega,\label{eq: NS incompressible}
\end{align}
where $\nu$ is the kinematic viscosity\footnote{We take the fluid density $\rho = 1$.}. On the boundary of any immersed body, we have the Dirichlet no-slip boundary condition:

\begin{equation}
    \vb{u}(\vb{x},t) = \vb{u}_b(\vb{x}, t), \quad \vb{x} \in \Gamma. \label{eq: NS bc}
\end{equation}
The imposed boundary velocity is subject to the volume-conservation constraint
\begin{equation}
    \int_{\Gamma} \vb{n} \cdot \vb{u}_b \dd S = 0, \label{eq: flux condition}
\end{equation}
where $\vb{n}(\vb{x}, t)$ is the unit normal on the boundary, and $\dd S$ is the length element on $\Gamma$. Throughout this work we consider rigid bodies with prescribed $\partial \vb{u}_b/ \partial t$ and $\vb{u}_b$, which includes static bodies and one-way coupled moving bodies.

\subsection{Discretization for static boundaries} \label{subsec: NS discretizations}
 To solve the governing equations, we use a low-storage Runge-Kutta (LSRK) projection method inspired by the approaches in~\cite{sanderse_accuracy_2012} and~\cite{vreman_projection_2014}. Each Runge-Kutta stage consists of three main steps, similar to a one-step projection approach: first, an intermediate velocity $\vb{u}^*$ is advanced without enforcing incompressibility; second, a divergence-correcting field for $\vb{u}^*$ is computed by solving a Poisson equation for a pseudo-pressure $\phi$; finally, the gradient of $\phi$ is subtracted from $\vb{u}^*$ to obtain a divergence-free velocity field.

In \cite{sanderse_accuracy_2012} it was shown that the pseudo-pressure computed in the first stage of such a Runge-Kutta based projection method can approximate the actual pressure field to high order temporal accuracy, but only under steady velocity boundary conditions. Here we adapt the algorithm to collocated grids, and extend it so that the pressure approximation has a high temporal order also under unsteady boundary conditions. To describe the resulting approach, we first consider the case where the computational domain is stationary, so that the surface normal vectors $\vb{n}$ remain constant in time; however, temporally varying velocity boundary conditions are allowed. The extension to moving boundaries is discussed subsequently in section~\ref{subsec: NS moving bc}.

Let the Runge-Kutta scheme contain $s$ stages for a single time step $\Delta t$. The scheme has coefficients  $a_{ij}$, where $i$ and $j$ represents different stages ($1 \leq i,j+1 \leq s$), along with the associated coefficients $c_i$, so that the time levels of the stages are $t^{(i)} = t^{(0)} + c_i \Delta t$. The intermediate velocity field, pseudo-pressure field, velocity field and velocity boundary condition at stage $i$ are represented as $\vb{u}^{*,(i)}$, $\phi^{(i)}$, $\vb{u}^{(i)}$, and $\vb{u}^{(i)}_b$, respectively. The initial velocity $\vb{u}^{(0)}$ is taken from the end of the previous time step. Moreover, we use a custom notation for the discrete differential gradient operators $\nabla^{D}$, $\nabla^{N}$, and $\nabla^{F}$, and similar superscripts for the Laplacian operator $\Delta$. These discrete operators encode the different IIM boundary treatments, corresponding to the use of, respectively, Dirichlet, Neumann and Free (unconstrained) boundary conditions in the construction of the uniformly weighted multivariate polynomial $p_c(\mathbf{x})$. Further, we use $\nabla^{AD}$ to denote the advection discretization explained in section~\ref{subsec:FD_IIM}: this discretization employs a Dirichlet condition at inflow boundaries and extrapolation at outflow boundaries, while using non-uniform decaying least squares weights for the polynomial construction.

With this notation, the proposed algorithm then applies the following three-step method for any stage $(i)$:

\textit{Step 1. Compute an intermediate velocity}
\begin{equation}
    \vb{u}^{*,(i)}= \vb{u}^{(0)} + \Delta t \sum_{j = 0}^{i-1} a_{ij}\left(-\vb{u}^{(j)} \cdot \nabla^{AD} \vb{u}^{(j)} + \nu \Delta^{D} \vb{u}^{(j)}\right)  \label{eq: step-1} 
\end{equation}

\textit{Step 2. Solve the Poisson equation}
\begin{equation}
    \Delta^{N} \phi^{(i)} = \frac{1}{c_i \Delta t} \nabla^{F} \cdot \vb{u}^{*,(i)}
    \label{eq: step-2}
\end{equation}

\textit{Step 3. Correct the velocity}
\begin{equation}
    \vb{u}^{(i)} = \vb{u}^{*,(i)} - c_i \Delta t \nabla^{F} \phi^{(i)} 
    \label{eq: step-3}
\end{equation}

In Step~1 the Dirichlet boundary conditions for $\vb{u}^{(j)}$ are prescribed as the no-slip condition given by $\vb{u}_b^{(j)}$. For Step~2 we formulate the Neumann boundary condition for the pseudo-pressure $\phi$ as

\begin{equation}
    \frac{\partial \phi}{\partial n}^{(i)} = \frac{1}{c_i \Delta t} \vb{n} \cdot \left( \vb{u}^{*,(i)} - \vb{u}_b^{(0)} - \Delta t \sum_{j = 0}^{i-1} a_{ij} \frac{\partial \vb{u}_b}{\partial t}^{(j)} \right), \quad \vb{x} \in \Gamma, 
    \label{eq: neumann bc}
\end{equation}
where $\vb{u}^{*,(i)}$ on the boundary is extrapolated using IIM, without using any boundary condition.

Conceptually, we note that the proposed scheme does not impose homogeneous Neumann boundary conditions on the pressure field; instead, it leaves the $\mathbf{u}^*$ field on the boundary unconstrained. This modification is consistently imposed in the algorithm: the right-hand side $\nabla \cdot \vb{u}^*$ in Step~2 is computed using \textit{extrapolated} values on the boundary, and $\mathbf{u}^*$ in the boundary condition~\eqref{eq: neumann bc} is extrapolated from the field. We will show below that this adaptation allows the pseudo-pressure to form a high-order estimate of the actual pressure, but introduces a discrete (convergent) error in the mass conservation. 

In the remainder of this subsection, we discuss the resulting algorithm respectively from the perspective of the normal surface velocity, the velocity divergence, and the overall accuracy of velocity and pressure. 
To aid the notation in the analysis, we define different types of spatial discretization schemes as $(\alpha, \beta)$, where $\alpha$ represents the order of the upwind finite difference discretization for the advection term, and $\beta$ is the order of the centered finite difference discretization for the other differential operators. For the fourth order scheme we use $(5,4)$, relying on the novel fifth-order upwind advection discretization presented in Section~\ref{sec: stabilityerror}. For comparisons, we also consider a third-order scheme $(3,4)$, and a second-order scheme $(3,2)$. The polynomial degree $k$ chosen for the IIM corrections is $\alpha + 1$ for the advection term, $\beta + 2$ for all other operations. As for time integration, we employ a third-order LSRK scheme for the $(5,4)$ and $(3,4)$ algorithms, and a second-order LSRK scheme for the $(3,2)$ algorithm.

\subsubsection{Normal surface velocity}
Here we demonstrate that $\vb{u}^{(i)}$ on the boundary remains consistent with $\vb{u}_b^{(i)}$. Taking the normal component of Equation~\eqref{eq: step-3} on the boundary and combining with Equation~\eqref{eq: neumann bc} yields
\begin{equation}
     \vb{n} \cdot \vb{u}^{(i)} = \vb{n} \cdot \left( \vb{u}^{*,(i)} - c_i \Delta t \nabla^F \phi^{(i)} \right) = \vb{n} \cdot \left(\vb{u}_b^{(0)} + \Delta t \sum_{j = 0}^{i-1} a_{ij}\frac{\partial \vb{u}_b}{\partial t}^{(j)}\right) + \mathcal{O}(\Delta t h^\beta), \quad \vb{x} \in \Gamma,
\end{equation}
where the mixed error term follows from $\vb{n} \cdot \nabla^{F}\phi^{(i)} = \partial \phi^{(i)}/ \partial n + \mathcal{O}(h^\beta)$ on the boundary. 
On the right-hand side, the term in parentheses represents a Runge-Kutta time integration for $\vb{u}_b^{(i)}$, whose accuracy order therefore follows the local accuracy of the Runge-Kutta method. Consequently,
\begin{equation}
    \vb{n} \cdot \vb{u}^{(i)} = \vb{n} \cdot \vb{u}_b^{(i)} + \mathcal{O}(\Delta t^{\gamma+1}) + \mathcal{O}(\Delta t h^\beta), \quad \vb{x} \in \Gamma. \label{eq: validate_flux}
\end{equation}
The mixed error term does not significantly affect accuracy because the CFL stability constraint implies $\Delta t \sim  h$, so that the mixed error converges as a high-order spatial error $\mathcal{O}(h^{\beta +1})$ under constant CFL spatio-temporal convergence. As a result, the no-through boundary condition on the updated velocity field is satisfied with a convergence order consistent with the spatial and temporal discretization schemes.

\subsubsection{Divergence} \label{subsec: divergence}
On collocated grids, it is challenging to discretely remove the divergence in the velocity predictor $\vb{u}^*$. Instead, we demonstrate here that the divergence converges with the same order of accuracy as the remainder of the scheme. 

We calculate here the divergence of the velocity as $\vartheta = \nabla^{F} \cdot \vb{u}$. 
To derive an expression for $\vartheta$, we apply the discrete divergence operator $\nabla^{F} \cdot$ to Equation~\eqref{eq: step-3}:

\begin{equation}
    \vartheta^{(i)} = \nabla^{F} \cdot \vb{u}^{*,(i)} - c_i \Delta t \nabla^{F}\cdot \nabla^{F} \phi^{(i)}.
    \label{eq: divergence condition}
\end{equation}
Using Equation~\eqref{eq: step-2} this can be rewritten as
\begin{equation}
    \vartheta^{(i)} = c_i \Delta t \left(\Delta^{N} \phi^{(i)} - \nabla^{F}\cdot \nabla^{F} \phi^{(i)} \right).
\end{equation}
Since both $\Delta^{N}$ and $\nabla^{F}\cdot \nabla^{F}$ are $\beta$th-order finite difference approximations of the Laplace operator, the term in parentheses converges as $\mathcal{O}(h^\beta)$. The asymptotic error in the divergence is thus $\vartheta^{(i)} = \mathcal{O}(\Delta t h^\beta)$. As a result, divergence of the updated velocity field appears as a mixed space-time error which, under a CFL stability condition, converges at an order $\beta + 1$. %

\subsubsection{Convergence order of pressure}\label{subsec: conv of p}
Here we demonstrate how the pseudo-pressure computed in the first stage, $\phi^{(1)}$, provides a $\gamma$th order approximation to the physical pressure at the beginning of the time step, $p^{(0)} = p^n$. Considering the first stage where $i = 1$ and substituting $\vb{u}^{*,(1)}$ from Step 1 into Step 2, one obtains
\begin{equation}
    \Delta^{N} \phi^{(1)} = \frac{1}{c_1 \Delta t} \nabla^{F} \cdot \vb{u}^{(0)} + \frac{a_{10}}{c_1}\nabla^{F} \cdot \left(-\vb{u}^{(0)} \cdot \nabla^{AD} \vb{u}^{(0)} + \nu \Delta^{D} \vb{u}^{(0)}\right),
    \label{eq: stage1 poisson}
\end{equation}
where $a_{10} = c_1$ holds for explicit Runge-Kutta methods. The first term on the right-hand side satisfies $\nabla^{F} \cdot \vb{u}^{(0)} = \vartheta^{(0)} = \mathcal{O}(\Delta t h^\beta)$; after multiplication by $1/(c_1 \Delta t)$ this yields an error of $\mathcal{O}(h^\beta)$. The second term on the right-hand side represents a high order discretization of a pressure Poisson equation. This pressure Poisson equation can be derived by taking the divergence of the Navier-Stokes momentum Equation~\eqref{eq: NS momentum} and applying the divergence-free condition Equation~\eqref{eq: NS incompressible}:
\begin{equation}
    \Delta p = - \nabla \cdot \left(\vb{u} \cdot \nabla \vb{u}\right) + \nu \nabla \cdot \Delta \vb{u}, \quad \vb{x} \in \Omega. \label{eq: p-field}
\end{equation}

As for the boundary conditions, the Neumann boundary condition for the pseudo-pressure $\phi^{(1)}$ is
\begin{equation}
    \frac{\partial \phi}{\partial n}^{(1)} = \frac{1}{c_1 \Delta t} \vb{n} \cdot \left( \vb{u}^{(0)} - \vb{u}_b^{(0)}\right) + \frac{a_{10}}{c_1}  \vb{n} \cdot 
 \left(-\frac{\partial \vb{u}_b^{(0)}}{\partial t} - \vb{u}^{(0)} \cdot \nabla^{AD} \vb{u}^{(0)} + \nu\Delta^{D} \vb{u}^{(0)}\right), \quad \vb{x} \in \Gamma.
 \label{eq: stage1 poissonbc}
\end{equation}
The first term on the right-hand side satisfies, using Equation~\eqref{eq: validate_flux}, the following estimate:
\begin{equation}
    \frac{1}{c_1 \Delta t} \vb{n} \cdot \left(\vb{u}^{(0)} - \vb{u}_b^{(0)}\right) = \mathcal{O}(\Delta t ^ \gamma) + \mathcal{O}(h ^ \beta).
\end{equation}
The second term on the right-hand side represents a discretization of a boundary condition for the pressure Poisson equation, obtained by projecting the momentum equation onto the normal direction using Equations~\eqref{eq: NS momentum} and~\eqref{eq: NS bc}:
\begin{equation}
    \frac{\partial p}{\partial n} = \vb{n} \cdot \nabla p = \vb{n} \cdot \left(-\frac{\partial \vb{u}_b}{\partial t} - \vb{u} \cdot \nabla \vb{u} + \nu\Delta \vb{u}\right), \quad \vb{x} \in \Gamma, \label{eq: p-bc}
\end{equation}

As a result, $\phi^{(1)}$ satisfies the pressure Poisson equation for $p^{(0)}$, and its error is dominated by the maximum value of the terms $\mathcal{O}(\Delta t^{\gamma}), \mathcal{O}(h^{\alpha})$ and $\mathcal{O}(h^{\beta})$. Thus, by taking $\vb{u}^{(0)}$ and $\phi^{(1)}$ as the solutions for $\vb{u}^{(0)}$ and $p^{(0)}$, respectively, a high order accurate velocity and pressure field can be obtained without additional computational cost.

We note that an alternative, more prevalent form of the Neumann boundary condition can be obtained by projecting Equation~\eqref{eq: step-3} onto the normal direction:
\begin{equation}
    \frac{\partial \phi}{\partial n}^{(i)} = \frac{1}{c_i \Delta t} \vb{n} \cdot\left(\vb{u}^{*,(i)} - \vb{u}_b^{(i)}\right), \quad \vb{x} \in \Gamma,\label{eq: traditional-p_bc}
\end{equation}
analogous to discussions in~\cite{veldman_missing_1990,Guy2005,vreman_projection_2014}. This condition enforces the velocity flux discretely, and matches the proposed boundary condition in Equation~\eqref{eq: neumann bc} when the boundary velocity is constant in time. For our algorithm, the accuracy can be verified by substituting Equation~\eqref{eq: step-1} into Equation~\eqref{eq: traditional-p_bc} for stage $i = 1$:
\begin{equation}
    \frac{\partial \phi}{\partial n}^{(1)} = \frac{1}{c_1 \Delta t}\vb{n} \cdot 
 \left(\vb{u}^{(0)} - \vb{u}_b^{(1)} \right) + \frac{a_{10}}{c_1}\vb{n} \cdot\left( -\vb{u}^{(0)} \cdot \nabla^{AD} \vb{u}^{(0)} + \nu\Delta^{D} \vb{u}^{(0)}\right), \quad \vb{x} \in \Gamma.
 \label{eq: stage1 poissonbc - classical}
\end{equation}
The key difference between Equation~\eqref{eq: stage1 poissonbc} and Equation~\eqref{eq: stage1 poissonbc - classical}, aside from error terms, is the term $(\vb{u}^{(0)} - \vb{u}_b^{(1)})/ (c_1 \Delta t)$. This term is a $\mathcal{O}(\Delta t)$ approximation of the boundary acceleration $\partial \vb{u}_b^{(0)}/ \partial t$ in Equation~\eqref{eq: p-bc}. Consequently,  using~\eqref{eq: traditional-p_bc} means that $\phi^{(1)}$ is a high-order approximation of $p^{(0)}$ only under steady velocity boundary condition, and reduces to an $\mathcal{O}(\Delta t)$ approximation of $p^{(0)}$ under unsteady boundary velocities. This is consistent with the analysis in~\cite{sanderse_accuracy_2012}. 

In section~\ref{subsec: free-space-bcs} below we compare the results of the two boundary conditions numerically and confirm this difference. 

For the external flows in non-periodic domains considered in Section~\ref{sec: resultsP2}, domain boundary conditions are required. These are chosen consistently as grid-aligned versions of the IIM boundary conditions. Specifically, we apply free-stream flows aligned with the $x$-direction, but allow steady and unsteady inflows. For these external flows, we apply free-slip conditions on the top and bottom domain boundaries. This yields the following domain boundary conditions:
\begin{equation}
    \begin{aligned}
    & \text{Inflow: }&u_x^{(i)} &= U_{\infty}^{(i)},&\quad u_y^{(i)} &= 0, &\quad  \frac{\partial \phi}{\partial x}^{(i)} &= \frac{1}{c_i\Delta t}\left( u_x^{*,(i)} - U_{\infty}^{(0)} - \Delta t \sum_{j=0}^{i-1} a_{ij} \frac{\partial U_{\infty}^{(j)}}{\partial t} \right),\\
    & \text{Outflow: }&\frac{\partial u_x}{\partial x}^{(i)} &= 0,&\quad \frac{\partial u_y^{(i)}}{\partial x} &= 0,&\quad  \frac{\partial \phi}{\partial x}^{(i)} &= 0 \\
    & \text{Free-slip: }&\frac{\partial u_x}{\partial y}^{(i)} &= 0,&\quad u_y^{(i)} &= 0, &\quad  \frac{\partial \phi}{\partial y}^{(i)} &= \frac{u_y^{*,(i)}}{c_i\Delta t},
\end{aligned}
\label{eq:farfield_bc}
\end{equation}
where $\vb{u} = (u_x, u_y)$ and $\vb{u}^* = (u_x^*, u_y^*)$, and $U_{\infty}^{(i)} = U_\infty(t^{(i)})$ is the imposed inflow velocity in the $x$-direction at stage $i$.

\subsection{Discretization for moving immersed boundaries} \label{subsec: NS moving bc}
For simulations involving moving boundaries, we combine the proposed projection method discussed in the previous section with the moving IIM method in Section~\ref{subsec: IIM moving}. At the beginning of the time step, we extend the velocity field $\vb{u}^{(0)}$ using the no-slip boundary condition into the body, so that $\vb{u}^{*,(0)} = E_D^{(0)}[\vb{u}^{(0)}]$. We then proceed with the time integration as follows:
\begin{align}
    & \vb{v}^{(i)} = \hat{a}_i\vb{v}^{(i-1)} + \Delta t E_F^{(i-1)}\left[-\vb{u}^{(i-1)} \cdot \nabla^{AD} \vb{u}^{(i-1)} + \nu \Delta^D \vb{u}^{(i-1)}\right]  \label{eq: moving integration a}\\
    & \vb{u}^{*,(i)} = \vb{u}^{*,(i-1)} + \hat{b}_{i}\vb{v}^{(i)} \label{eq: moving integration b}\\
    & t^{(i)} = t^{(0)} + c_i\Delta t,
\end{align}
where $\vb{v}$ is a temporary field storing the history values in low-storage RK integrators. The coefficients $a_{i,j}$ for any low-storage Runge-Kutta method can be re-organized to $\hat{a}_i$ and $\hat{b}_i$ as in Equation~\eqref{eq: moving integration a} and~\eqref{eq: moving integration b}~\cite{williamson_low-storage_1980}, requiring only the storage of the intermediate velocity $\vb{u}^*$ and one history field $\vb{v}$, with $\hat{a}_1$ always be zero. We note that equations~\eqref{eq: moving integration a} and \eqref{eq: moving integration b} integrate the temporary field $\vb{v}$ and intermediate velocity field $\vb{u}^*$ within the flow domain, as well as one layer of grid points extended into the body.

Next, let $\vb{n}^{(i)}$ represent surface normal vectors of the body at stage $i$, as defined by $\Gamma^{(i)}$. We then solve the Poisson system for $\phi^{(i)}$:
\begin{align}
    &\Delta^{N} \phi^{(i)} =  \frac{1}{c_i \Delta t} \nabla^{F} \cdot \vb{u}^{*,(i)} - \frac{1}{\Tilde{c}_i \Delta t} \nabla^F \cdot \vb{u}^{(0)}, \label{eq: poisson moving}\\
    &\frac{\partial \phi}{\partial n}^{(i)} = \frac{1}{c_i \Delta t} \vb{n}^{(i-1)} \cdot \left( \vb{u}^{*,(i)} - \vb{u}_b^{(0)} - \Delta t \sum_{j = 0}^{i-1} a_{ij} \frac{\partial \vb{u}_b}{\partial t}^{(j)} \right) - \frac{1}{\Tilde{c}_i \Delta t} \vb{n}^{(i-1)} \cdot \left( \vb{u}^{(0)} - \vb{u}_b^{(0)}\right) , \quad \vb{x} \in \Gamma^{(i-1)},  \label{eq: p-bc moving}
\end{align}
where $\vb{u}^{(0)}$ on the boundary is extrapolated using IIM without boundary condition. This system is only solved in the flow domain outside $\Gamma^{(i-1)}$, ignoring the extended values of $\vb{u}^{*,(i)}$. 
Finally, we correct the velocity given $\phi^{(i)}$ and move the boundary by zeroing the velocity field inside $\Gamma^{(i)}$:

\begin{equation}
    \vb{u}^{(i)} = Z^{(i)}\left[\hat{\vb{u}}^{*,(i)} - c_i \Delta t E_F^{(i-1)}\left[\nabla^F \phi^{(i)}\right]\right], \label{eq: moving-last}
\end{equation}

Compared to the stationary boundary case defined in Equations~\eqref{eq: step-2} and~\eqref{eq: neumann bc}, the above equations~\eqref{eq: poisson moving} and \eqref{eq: p-bc moving} each subtract a new term with pre-factor $1/\Tilde{c}_i$. The coefficient $\Tilde{c}_i$ is introduced specifically for moving boundary simulations to control some of the numerical noise in the first-stage pressure signal associated with the discrete transitions in intersection locations between (sub)-steps of the time integration. Specifically, when $\Tilde{c}_i \to \infty$, the Poisson system is the same as the algorithm described in section~\ref{subsec: NS discretizations} for stationary boundary simulations. This choice is used for all stages except the first one where $i = 1$, where we wish to use $\phi^{(1)}$ as a high-order estimate for $p^{(0)}$. For finite values of $\Tilde{c}_i$ in the first stage we can assess the modification by substituting $\vb{u}^{*,(1)}$ into Equation~\eqref{eq: poisson moving}:
\begin{equation}
    \Delta^N \phi^{(1)} = \frac{1}{c_1 \Delta t} \left(1 - \frac{c_1}{\Tilde{c}_1}\right) \nabla^F \cdot \vb{u}^{(0)} + \nabla^F \cdot \left(-\vb{u}^{(0)} \cdot \nabla^{AD} \vb{u}^{(0)} + \nu \Delta^D \vb{u}^{(0)}\right). \label{eq: modified-poisson}
\end{equation}
The corresponding Neumann boundary condition is
\begin{equation}
     \frac{\partial \phi}{\partial n}^{(1)} = \frac{1}{c_1 \Delta t} \left(1 - \frac{c_1}{\Tilde{c}_1}\right) \vb{n}^{(0)} \cdot \left( \vb{u}^{(0)} - \vb{u}_b^{(0)}\right) +  \vb{n}^{(0)} \cdot \left(-\frac{\partial \vb{u}_b^{(0)}}{\partial t} - \vb{u}^{(0)} \cdot \nabla^{AD} \vb{u}^{(0)} + \nu\Delta^D \vb{u}^{(0)}\right), \quad \vb{x} \in \Gamma^{(0)},
     \label{eq: modified-poisson-bc}
\end{equation}

We see that at the first stage, $\Tilde{c}_1$ represents a parameter that allows decoupling the 'solenoidal-projection' part of the right-hand side of the pseudo-pressure Poisson problem, from the 'pressure projection' part. For instance, setting $\Tilde{c}_1 = c_1$ eliminates the pressure projection, whereas $\Tilde{c}_1 \to \infty$ recovers the projection algorithm proposed in section~\ref{subsec: NS discretizations}. This makes $\Tilde{c}_1$ a useful parameter to tune the time-varying numerical fluctuations in the pressure field that originate from $\nabla^F \cdot \mathbf{u}^{(0)}$, and their effect on $\phi^{(1)}$ as an estimate of $p^{(0)}$. However, choosing a finite value of $\Tilde{c}_1$ comes at the cost of a divergence error: substituting Equation~\eqref{eq: poisson moving} into Equation~\eqref{eq: divergence condition} yields an updated estimate for $\vartheta$:
\begin{equation}
    \vartheta^{(1)} = c_1 \Delta t \left(\Delta^{N} \phi^{(1)} - \nabla^{F}\cdot \nabla^{F} \phi^{(1)} \right) + \frac{c_1}{\Tilde{c}_1}\vartheta^{(0)},
\end{equation}
which demonstrates that an additional divergence error at the end of the first stage is added to $\vartheta^{(i)}$ whenever $\Tilde{c}_1$ is finite. However, by using $\Tilde{c}_1 \to \infty$ in all subsequent stages, this error in practice does not affect simulation results. Section~\ref{subsec: results p1 moving} discusses the effect of varying $\Tilde{c}_1$ in numerical simulations.

\subsection{Implementation and solution method}
\label{subsec:implementation}
All methods in this work are implemented in Julia~\cite{Julia-2017} and are run in serial on a modern laptop or workstation. In most practical flows we impose Neumann boundary conditions on immersed boundaries, and Neumann or periodic conditions on domain boundaries. Consequently, the associated linear system is singular, as the pressure field is only determined up to a constant. Further, due to discretization  errors the discrete right-hand side may not lie in the range of the system matrix. To resolve these issues, we follow the method proposed by \cite[Chapter~5.6.4]{Trottenberg2000} and \cite{gabbard_high_2025} to augment the Poisson system. Writing Equation~\eqref{eq: step-2} in matrix form $\mathcal{L} \Phi = \vb{f}$, where $\Phi$ is the vector of $\phi$ values at grid points in $\Omega$, and $\mathbf{f}$ represents the right-hand side, we construct:
\begin{equation}
    \begin{bmatrix}
        \mathcal{L} & \vb{r} \\
        \vb{r}' & 0
    \end{bmatrix}
    \begin{bmatrix}
        \Phi \\
        \lambda
    \end{bmatrix}
    =  \begin{bmatrix}
        \vb{f} \\
        \mu
    \end{bmatrix},
    \label{eq:augmented_poisson}
\end{equation}
where $\mathbf{r} = [1,1, \dots, 1]^T$ is the right null vector of $\mathcal{L}$, and $\lambda$ and $\mu$ are scalars. This system is full rank and is solved using a sparse direct solver. The solution $\Phi$ satisfies:
\begin{align}
    \mathcal{L} \Phi = \vb{f} - \lambda\vb{r}, \label{eq: sigularity_beta} \\
    \vb{r}' \Phi = \mu. \label{eq: singularity_eta}
\end{align}
Equation~\eqref{eq: sigularity_beta} indicates that $\lambda$ captures the discretization error in the compatibility condition, while Equation~\eqref{eq: singularity_eta} illustrates that $\mu$ controls the sum of $\phi$ over domain grid points. Throughout this work we set $\mu = 0$ unless otherwise specified.

\section{Results part 1: convergence with exact solutions}
\label{sec: resultsP1}
In this section, we verify the proposed Navier-Stokes discretization for simulations with no boundaries, static boundaries, and moving boundaries respectively. Additionally, we assess the impact of the first-stage modification in reducing numerical noise in simulations involving moving boundaries. Throughout this section we use the Taylor-Green vortex field on a unit square domain, for which the exact solution to the 2D Navier-Stokes is given by:
\begin{align}
    & u_x(\vb{x},t) = \cos(2\pi x)\sin(2\pi y) e^{-8\pi^2\nu t},\label{eq: TG-1}\\
    & u_y(\vb{x},t) = -\sin(2\pi x)\cos(2\pi y) e^{-8\pi^2\nu t}, \label{eq: TG-2}\\
    & p(\vb{x},t) = -\left(\cos(4\pi x) + \cos(4\pi y)\right) e^{-16\pi^2\nu t}/4, \label{eq:TG-3}
\end{align}
with $\vb{x} = (x,y)$ and $\vb{u} = (u_x,u_y)$. For purposes of convergence, each simulation presented in this section is run at a constant time step $\Delta t$.

\subsection{Free-space} \label{subsec: free-space}
\begin{figure}
    \begin{subfigure}[t]{0.3\linewidth}
        \centering
        \includegraphics[width=0.67\textwidth]{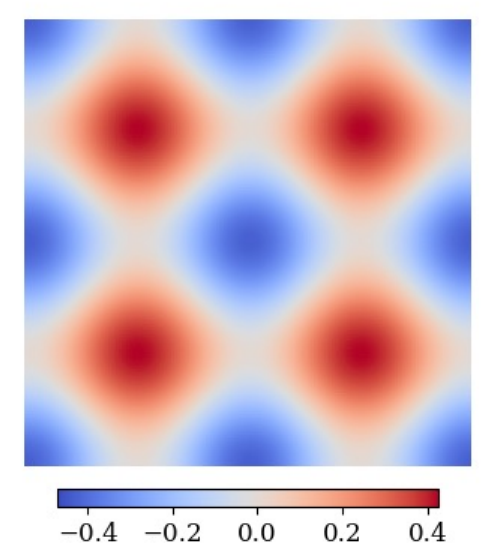}
        \caption{Pressure field at $t=1$ ($N=128$, $(3,4)$ algorithm).}
        \label{subfig: freespace illu}
    \end{subfigure}
    \hspace{0.02\textwidth}
    \begin{subfigure}[t]{0.31\linewidth}
        \centering
        \includegraphics[width=\textwidth]{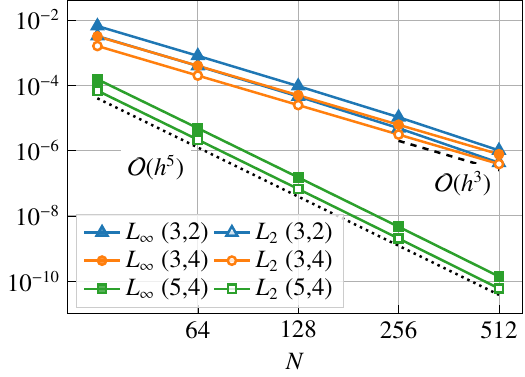}
        \caption{Convergence of $u_x$ at $t=1$ compared with the exact solution ($CFL=0.2$).}
        \label{subfig: freespace u CFL}
    \end{subfigure}
    \hspace{0.02\textwidth}    
    \begin{subfigure}[t]{0.30\linewidth}
        \centering
        \includegraphics[width=\textwidth]{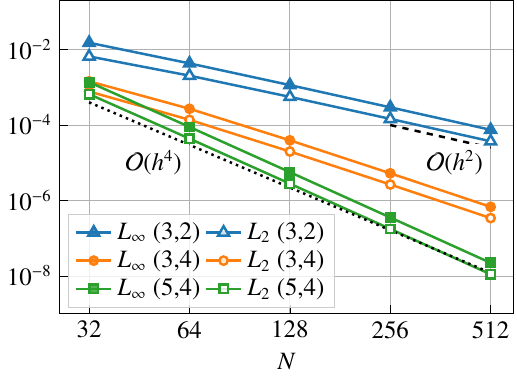}
        \caption{Convergence of $p$ at $t=1$ compared with the exact solution ($CFL=0.2$).}
        \label{subfig: freespace p CFL}
    \end{subfigure}
    \vspace{0.02\textwidth}    
    \begin{subfigure}[t]{0.30\linewidth}
        \centering
        \includegraphics[width=\textwidth]{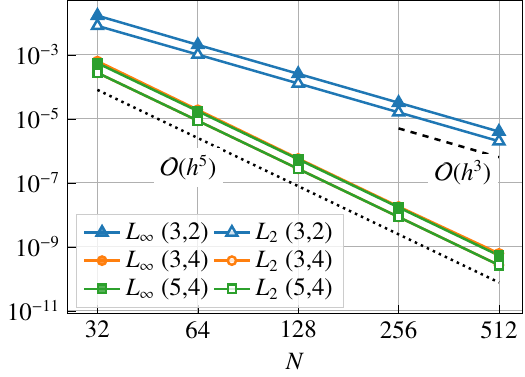}
        \caption{Convergence of $\vartheta$ at $t=1$ compared with the exact solution $\vartheta = 0$ ($CFL=0.2$).}
         \label{subfig: freespace div CFL}
    \end{subfigure}
    \hspace{0.02\textwidth}
    \begin{subfigure}[t]{0.32\linewidth}
        \centering
        \includegraphics[width=\textwidth]{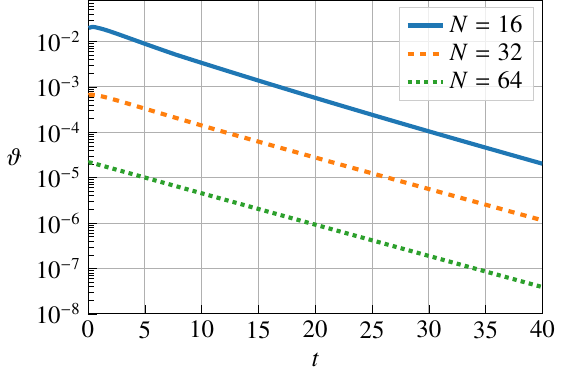}
        \caption{Time evolution of the $L_\infty$ error in $\vartheta$ at different resolutions ($CFL=0.2$, $(3,4)$ algorithm).}
        \label{subfig: free-div-history}
    \end{subfigure}
    \hspace{0.02\textwidth}
    \begin{subfigure}[t]{0.305\linewidth}
        \centering
        \includegraphics[width=\textwidth]{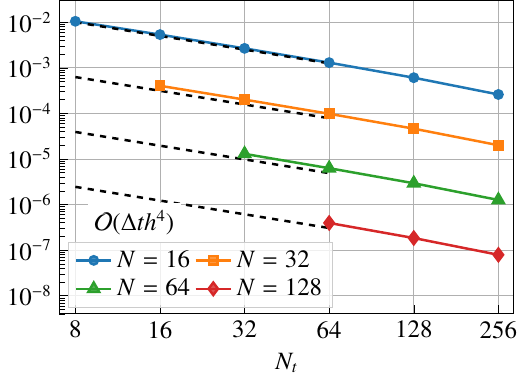}
        \caption{Temporal convergence of $u_x$ at $t=0.5$ in the $L_\infty$ norm, compared to solutions computed with $N_t = 512$ ($(3,4)$ algorithm).}
        \label{subfig: time-conv-u}
    \end{subfigure}
    \caption{Illustration and convergence plots of the free-space Taylor-Green vortex field.} 
    \label{fig: freespace error}
\end{figure}

We first report convergence results for simulations of the Taylor-Green vortex in a two-dimensional periodic domain $[0,1) \times [0, 1)$ without any immersed boundaries, discretized using an $N \times N$ grid. Here the viscosity is set to $\nu = 0.001$ so that the Reynolds number $Re = 1/\nu = 1000$. All simulations are initialized from Equations~\eqref{eq: TG-1} and \eqref{eq: TG-2} at $t = 0$ and are evolved using the algorithm presented in section~\ref{subsec: NS discretizations}. The total number of time steps in any simulation is defined as $N_t = t / \Delta t$. The computed pressure field at $t = 1$ is illustrated in Figure~\ref{subfig: freespace illu}. 

Figures~\ref{subfig: freespace u CFL}, ~\ref{subfig: freespace p CFL} and~\ref{subfig: freespace div CFL}  show convergence of the $x$-velocity component, pressure, and divergence respectively. The convergence tests are conducted at $\Delta t = 0.2 h$ with $h=1/N$, corresponding to a CFL number based on the initial maximum velocity of $0.2$. For each quantity and scheme we report the $L_2$ and $L_\infty$ norms of the errors compared to the exact solution at $t = 1$. Since the viscosity is relatively small, errors for the velocity fields are dominated by the advection term. Hence, Figure~\ref{subfig: freespace u CFL} shows third-order convergence for the $(3,2)$ and $(3,4)$ schemes, and fifth order for the $(5,4)$ scheme in both the $L_2$ and $L_\infty$ norm. The pressure in Figure~\ref{subfig: freespace p CFL} converges consistently with the minimum spatial discretization order, which is second for the $(3,2)$ scheme, third for the $(3,4)$ scheme and fourth for the $(5,4)$ scheme. Moreover, the fixed CFL convergence of divergence in Figure~\ref{subfig: freespace div CFL} confirms the mixed error of $\mathcal{O}(\Delta t h^\beta)$, which here appears as a $\mathcal{O}(h^{\beta+1})$ convergence order. Figure~\ref{subfig: free-div-history} shows the evolution of the $L_\infty$ norm of the divergence field at different resolutions up to $t=40$, demonstrating that the projection method remains stable during the simulation.

Finally, Figure~\ref{subfig: time-conv-u} shows the temporal convergence of the velocity component $u_x$ at $t=0.5$. The error is computed in the $L_\infty$ norm with respect to the discrete solution obtained with $N_t=512$ for each resolution, so that we can analyze the temporal error in isolation from spatial discretization errors.
The results confirm the presence of the mixed error term $\mathcal{O}(\Delta t h^\beta)$ in the velocity, which can be observed in pressure and divergence as well (not shown here). However, as argued in section~\ref{subsec: IIM moving} and evident from panels (b)--(d) in Figure~\ref{fig: freespace error}, this error does not dominate in practical applications.

\subsection{Comparison of pressure boundary conditions on the domain boundaries}
\label{subsec: free-space-bcs}
To highlight the difference between the proposed boundary condition~\eqref{eq: neumann bc} and the alternative boundary condition~\eqref{eq: traditional-p_bc}, we reconsider a free-space simulation of the Taylor-Green vortex. As opposed to the previous subsection, the simulations are conducted in the shifted two-dimensional unit square $[1/7,8/7) \times [1/7, 8/7)$, where the shift is applied to ensure non-zero velocity gradients at the boundaries. The left and right boundaries of the domain are now considered domain boundaries where Dirichlet boundary conditions on velocity and Neumann boundary conditions on the pseudo-pressure are imposed. The top and bottom domain boundaries are still considered periodic, as in the previous subsection. For this case we set the viscosity $\nu = 0.01$, so that the spatial discretization errors are reduced and the temporal error becomes dominant.

\begin{figure}
    \begin{subfigure}[t]{0.45\linewidth}
        \centering
        \includegraphics[width=0.62\textwidth]{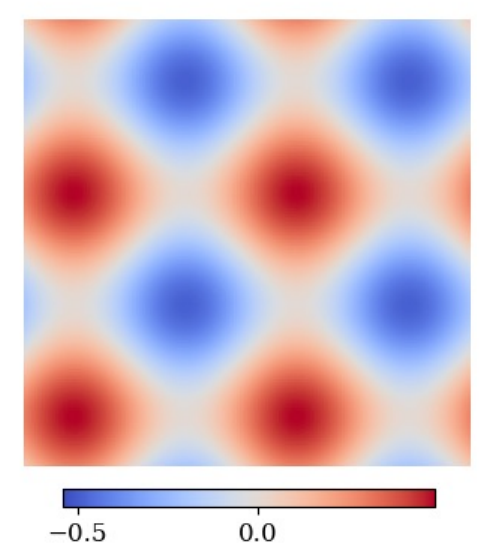}
        \caption{Pressure field at $t=0.01$ ($N=128$, $(3,4)$ algorithm).}
        \label{subfig: domain BC illu - BC test}
    \end{subfigure}
    \hspace{0.02\textwidth}
    \begin{subfigure}[t]{0.42\linewidth}
        \centering
        \includegraphics[width=\textwidth]{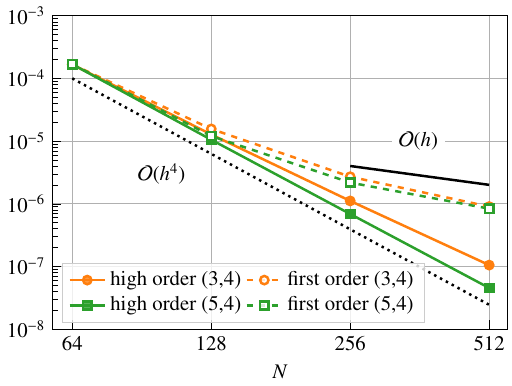}
        \caption{Convergence of the $L_\infty$ error in $p$ at $t=0.01$, compared to the exact solution ($CFL = 0.04$).}
        \label{subfig: Neumann BC comparison}
    \end{subfigure}
    
    \caption{Illustration and convergence plots of the Taylor-Green vortex with different Neumann boundary conditions enforced on the domain, where `high-order' refers to the high order Neumann BC~\eqref{eq: neumann bc}, and `first-order' refers to the first-order Neumann BC~\eqref{eq: traditional-p_bc}. $(\alpha, \beta)$ represents using $\alpha$th-order scheme for advection term and $\beta$th-order schemes for others. LSRK3 is used for both $(3,4)$ and $(5,4)$.} 
    \label{fig: test Neumann BC}
\end{figure}

Figure~\ref{subfig: domain BC illu - BC test} shows the pressure field at $t = 0.01$, while Figure~\ref{subfig: Neumann BC comparison} presents the convergence results for the three schemes with $\Delta t = 0.04 h$. This time step is chosen to keep the maximum Fourier number across all resolutions below $0.2$ for stability. Here `high-order' refers to the proposed high-order Neumann boundary condition \eqref{eq: neumann bc}, and `first-order' refers to the more prevalent first-order version \eqref{eq: traditional-p_bc}. The results show that for high-order schemes $(5,4)$ and $(3,4)$, the proposed Neumann condition~\eqref{eq: neumann bc} maintains high-order convergence as the resolution increases. In contrast, the convergence with the first-order boundary condition~\eqref{eq: traditional-p_bc} stalls to first order as resolution increases. Since our convergence is performed with $\Delta t \sim h$, this first-order slope reflects the expected $\mathcal{O}(\Delta t)$ error discussed in section~\ref{subsec: conv of p}.

\subsection{Static immersed boundaries}
\begin{figure}
\centering
    \begin{subfigure}[t]{0.3\linewidth}
        \centering
        \includegraphics[width=0.65\textwidth]{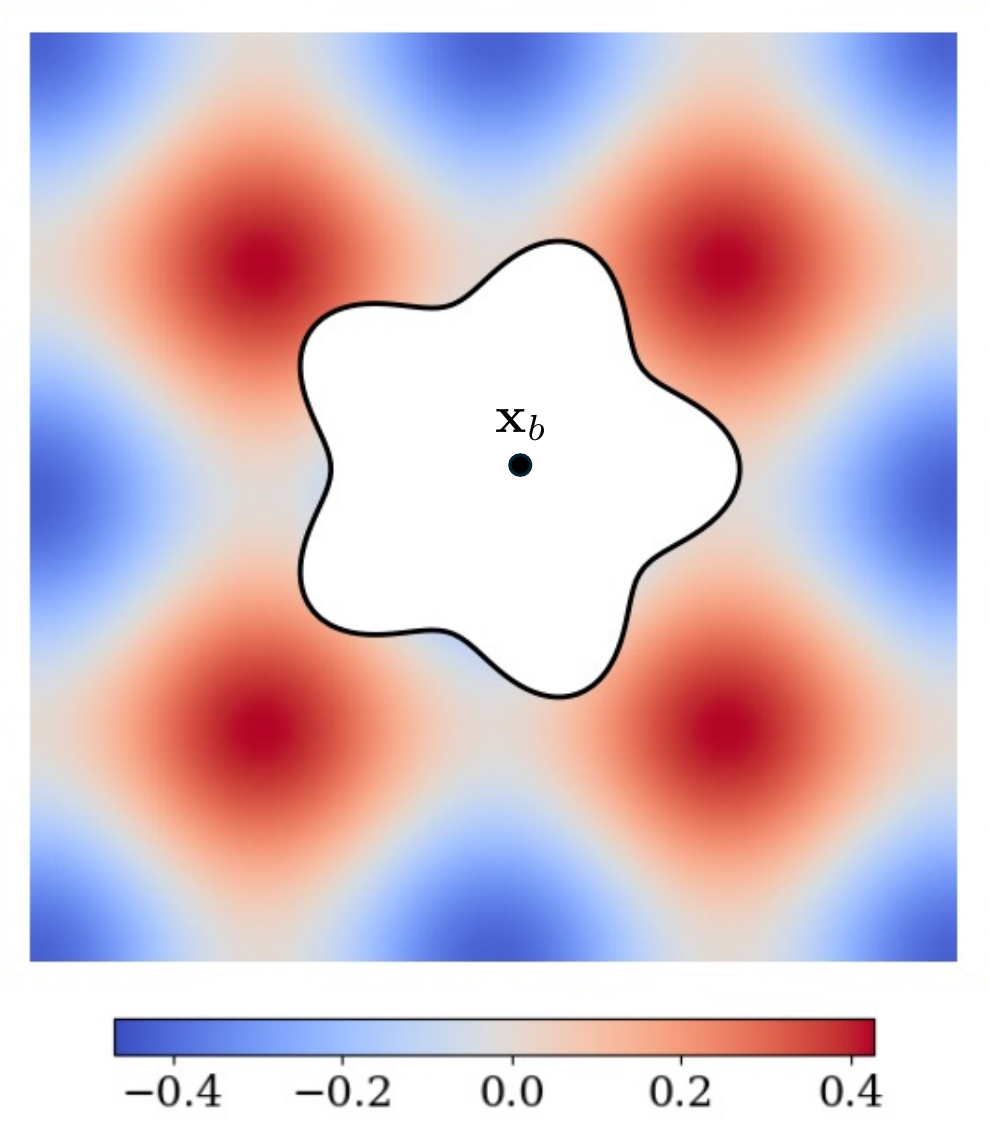}
        \caption{Pressure field at $t=1$ with an embedded star centered at $\vb{x}_b$ ($N=128$, $(3,4)$ algorithm).}
        \label{subfig: static star illu}
    \end{subfigure}
        \hspace{0.02\textwidth}
    \begin{subfigure}[t]{0.305\linewidth}
        \centering
        \includegraphics[width=\textwidth]{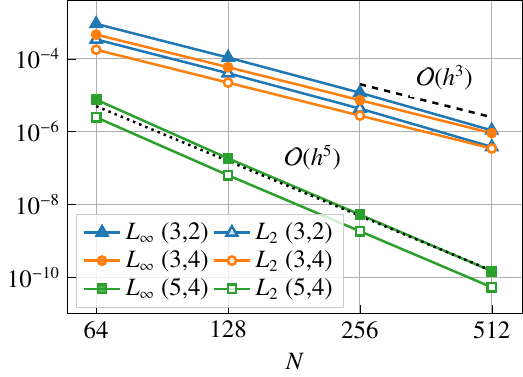}
        \caption{Convergence of $u_x$ at $t=1$ compared with the exact solution ($CFL=0.2$).}
        \label{subfig: static star CFL u}
    \end{subfigure}
    \hspace{0.02\textwidth}
    \begin{subfigure}[t]{0.31\linewidth}
        \centering
        \includegraphics[width=\textwidth]{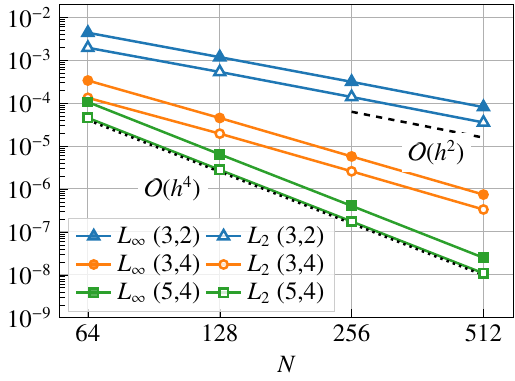}
        \caption{Convergence of $p$ at $t=1$ compared with the exact solution ($CFL=0.2$).}
        \label{subfig: static star CFL p}
    \end{subfigure}

    \vspace{0.02\textwidth}

    \begin{subfigure}[t]{0.30\linewidth}
        \centering
        \includegraphics[width=\textwidth]{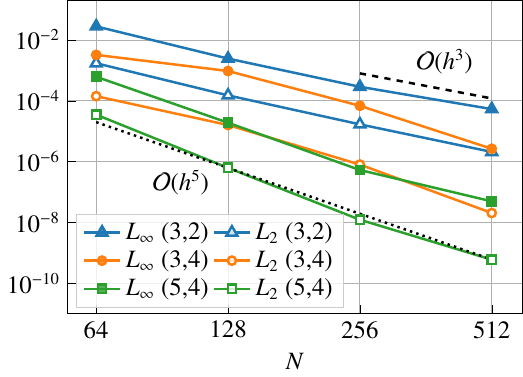}
        \caption{Convergence of $\vartheta$ at $t=1$ compared with the exact solution $\vartheta = 0$ ($CFL=0.2$).}
        \label{subfig: static star CFL div}
    \end{subfigure}
    \hspace{0.02\textwidth}
    \begin{subfigure}[t]{0.31\linewidth}
        \centering
        \includegraphics[width=\textwidth]{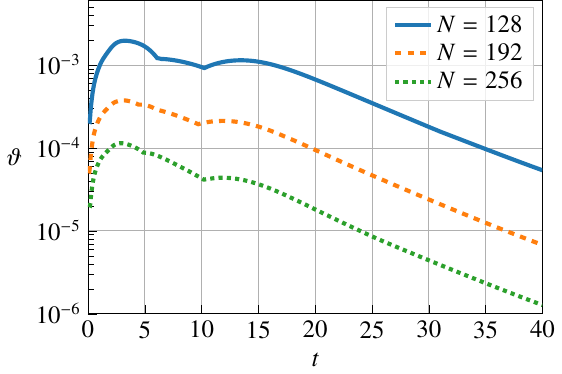}
        \caption{Time evolution of the $L_\infty$ error in $\vartheta$ at different resolutions ($CFL=0.2$, $(3,4)$ algorithm).}
        \label{subfig: static star long-time divergence}
    \end{subfigure}
    \hspace{0.02\textwidth}
    \begin{subfigure}[t]{0.30\linewidth}
        \centering
        \includegraphics[width=\textwidth]{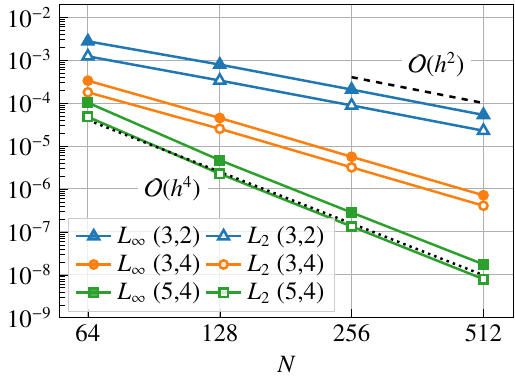}
        \caption{Convergence of the surface pressure at $t=1$ compared with the exact solution ($CFL=0.2$).}
         \label{subfig: static star CFL pdis}
    \end{subfigure}    
    \caption{Illustration and convergence plots of the Taylor-Green velocity field evolved with a stationary immersed star-shaped body.} 
    \label{fig: static_star}
\end{figure}

We use the 2D Taylor-Green vortex field again to verify the proposed Navier-Stokes algorithm applied to flows with embedded bodies. To test a geometry with both convex and concave boundaries, we consider a star-shaped obstacle embedded within the $[0, 1) \times [0, 1)$ fluid domain. Its shape is given by Equation~\eqref{eq: star-like levelset} with $\vb{x}_b = (0.51, 0.53)$, $r_b = 0.22$, $r_d = 0.035$, and $N_s = 5$, as shown in Figure~\ref{subfig: static star illu}. All simulations are initialized with the exact Taylor-Green vortex solution at $t = 0$ and run until $t = 1$ with $\Delta t = 0.2 h$. The exact velocity and acceleration fields are used to evaluate the boundary terms $\vb{u}_b$ and $(\partial \vb{u}/ \partial t)_b$ respectively, while the exact mean of $p$ is applied to the augmented Poisson system~\eqref{eq:augmented_poisson} as $\mu$. With this setup, the flow field outside the star should remain equal to the exact, free-space Taylor-Green vortex, making this a convenient setup to test convergence of the immersed boundary treatment.

The $L_\infty$ and $L_2$ norms of the error fields for $u_x$, $p$ and $\vartheta$ at $t=1$ are calculated for different resolutions $N$ and the schemes $(5,4)$, $(3,4)$, and $(3,2)$, as shown in Figure~\ref{subfig: static star CFL u}, ~\ref{subfig: static star CFL p} and~\ref{subfig: static star CFL div} respectively. The plots confirm that the convergence orders of all the quantities across the different schemes are as expected. Moreover, comparing Figure~\ref{subfig: static star CFL div} with the free-space result in Figure~\ref{subfig: freespace div CFL}, we find that the presence of the embedded boundary increases the magnitude of the divergence error, but the error still converges with high order. Figure~\ref{subfig: static star long-time divergence} illustrates the long-time evolution of the velocity divergence $\vartheta$, confirming that the simulation remains stable over time. 
Finally, we examine the accuracy of the computed surface pressure distribution, which is crucial for practical applications involving force calculations. The surface pressure is computed at the control points using polynomial extrapolation based on our IIM least-squares polynomial. The error convergence plot of the surface pressure distribution on the star is shown in Figure~\ref{subfig: static star CFL pdis}, demonstrating that the convergence orders and error values of the different algorithms closely reflect the results of the pressure field shown in Figure~\ref{subfig: static star CFL p}.

\subsection{Moving immersed boundaries}
\label{subsec: results p1 moving}
\begin{figure}
\centering
    \begin{subfigure}[t]{0.3\linewidth}
        \centering
        \includegraphics[width=0.65\textwidth]{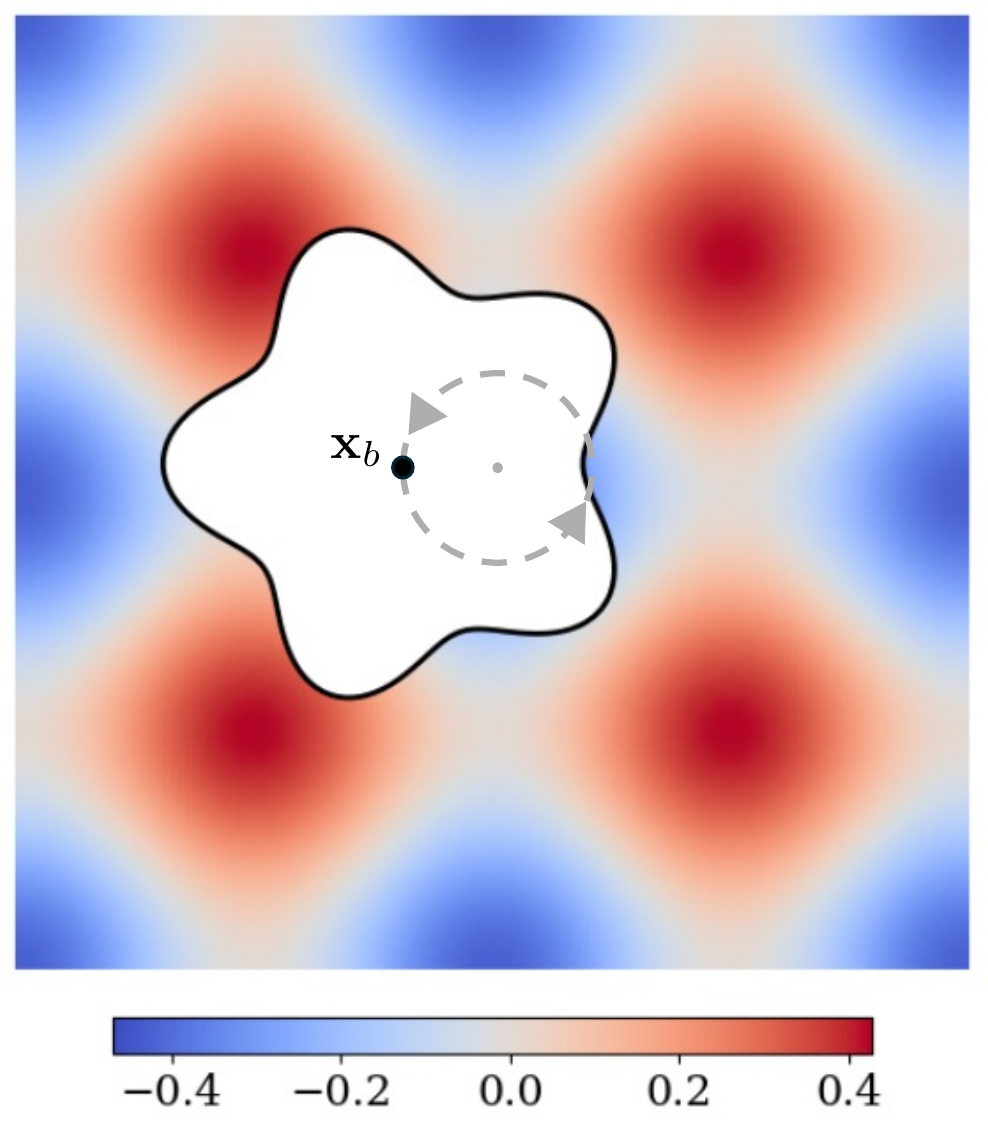}
        \caption[]{Pressure field at $t=1$ with an embedded star moving and rotating along the dashed gray counter-clockwise circle ($N=128$, $(3,4)$ algorithm).}
        \label{subfig: rotating star illu}
    \end{subfigure}
    \hspace{0.02\textwidth}
    \begin{subfigure}[t]{0.305\linewidth}
        \centering
        \includegraphics[width=\textwidth]{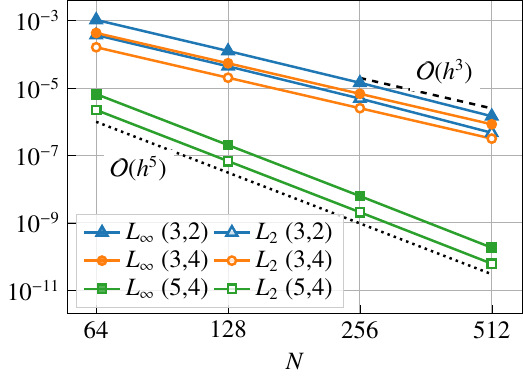}
        \caption{Convergence of $u_x$ at $t=1$ compared with the exact solution ($CFL=0.2$).}
        \label{subfig: rotating star CFL u}
    \end{subfigure}
    \hspace{0.02\textwidth}
    \begin{subfigure}[t]{0.31\linewidth}
        \centering
        \includegraphics[width=\textwidth]{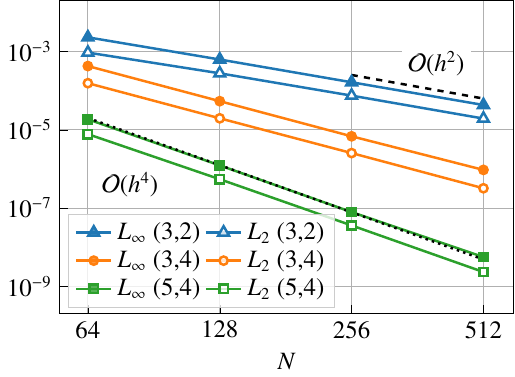}
        \caption{Convergence of $p$ at $t=1$ compared with the exact solution ($CFL=0.2$).}
        \label{subfig: rotating star CFL p}
    \end{subfigure}

    \vspace{0.02\textwidth}

    \begin{subfigure}[t]{0.30\linewidth}
        \centering
        \includegraphics[width=\textwidth]{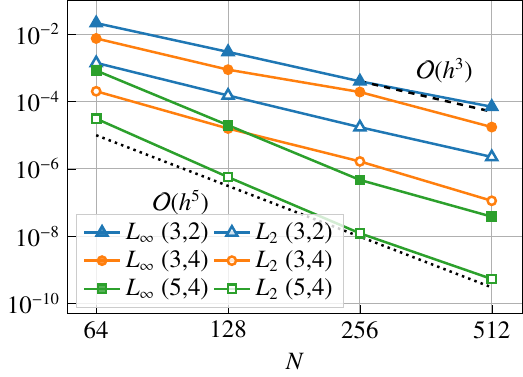}
        \caption{Convergence of $\vartheta$ at $t=1$ compared with the exact solution $\vartheta = 0$ ($CFL=0.2$).}
        \label{subfig: rotating star CFL div}
    \end{subfigure}
    \hspace{0.02\textwidth}
    \begin{subfigure}[t]{0.31\linewidth}
        \centering
        \includegraphics[width=\textwidth]{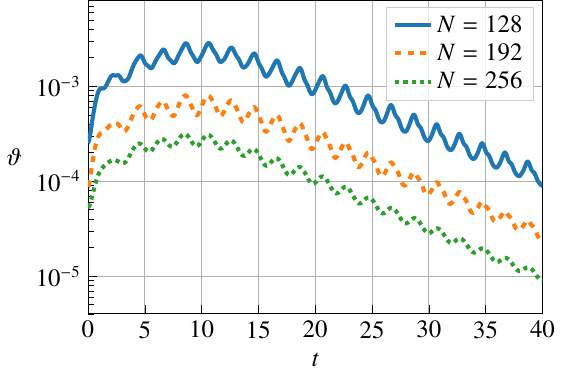}
        \caption{Time evolution of the $L_\infty$ error in $\vartheta$ at different resolutions ($CFL=0.2$, $(3,4)$ algorithm).}
        \label{subfig: rotating star long-time divergence}
    \end{subfigure}
    \hspace{0.02\textwidth}
    \begin{subfigure}[t]{0.30\linewidth}
        \centering
        \includegraphics[width=\textwidth]{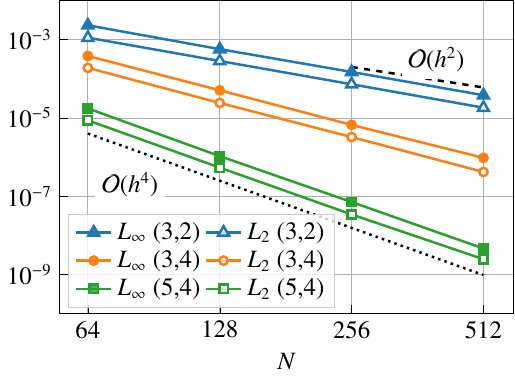}
        \caption{Convergence of the surface pressure at $t=1$ compared with the exact solution ($CFL=0.2$).}
         \label{subfig: rotating star CFL pdis}
    \end{subfigure}
    \caption{Illustration and convergence plots of the Taylor-Green velocity field evolved with a rotating immersed star-shaped body.} 
    \label{fig: rotating_star}
\end{figure}

Lastly, we extend the convergence results to a case where the star-shaped obstacle undergoes a prescribed motion within the Taylor-Green vortex field. Specifically, the star undergoes both rotation about its center $\vb{x}_b$ and counter-clockwise orbital motion around the point $(0.51, 0.53)$, as shown in Figure~\ref{subfig: rotating star illu}. The rotations are prescribed with angular velocity of $\Omega = \pi$, and a full rotation occurs every $t = 2$ units. The star is initially positioned at $\vb{x}_b = (0.61, 0.53)$.

\begin{figure}
    \begin{subfigure}[t]{0.45\linewidth}
        \centering
        \includegraphics[width=\textwidth]{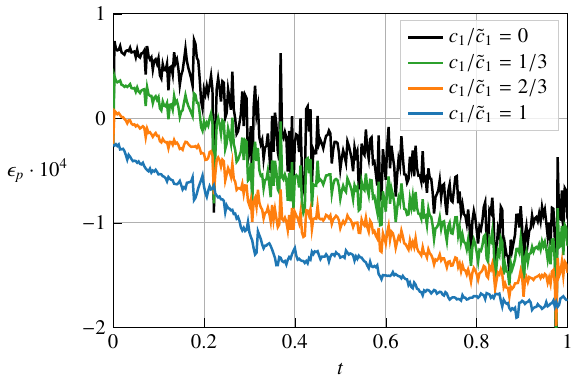}
        \caption{Time evolution of the pressure error $\epsilon_p$.}
        \label{subfig: cm history}
    \end{subfigure}
    \hspace{0.02\textwidth}
    \begin{subfigure}[t]{0.49\linewidth}
        \centering
        \includegraphics[width=\textwidth]{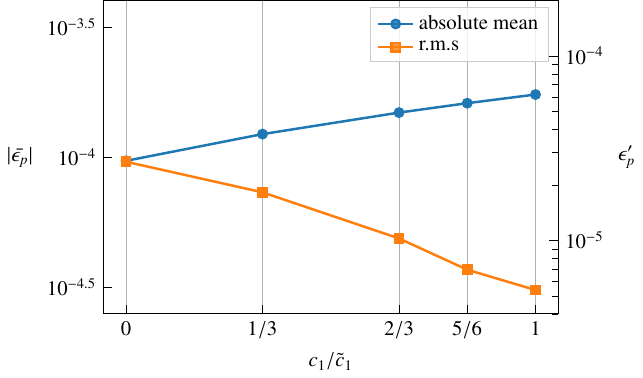}
        \caption{Absolute mean  $| \Bar{\epsilon}_p |$ and r.m.s. $\epsilon_p '$ between $t = 0.7$ and $t = 1.0$.}
        \label{subfig: cm analysis}
    \end{subfigure}
    
    \caption{Influence of $c_1/\Tilde{c}_1$ on the accuracy and fluctuations in the pressure field computed using linear resolution $N=64$ at location $(0.8, 0.8)$ in the domain. As $\Tilde{c}_1$ decreases, the oscillations in the error reduce whereas the mean error increases.} 
    \label{fig: cm effect}
\end{figure}

We first assess the impact of the first-stage modification constant $\Tilde{c}_1$ on the temporal pressure variations. To do so, we plot the pressure error history $\epsilon_p = p - p_{exact}$ at the point $(0.8, 0.8)$ for different values of $c_1/\Tilde{c}_1$ in Figure~\ref{subfig: cm history}. Here $p_{exact}$ denotes the exact pressure solution, and the linear resolution $N$ equals $N = 64$. This spatial resolution is relatively low, which is chosen here to exaggerate the magnitude of the numerical variations compared to better resolved simulations. Figure~\ref{subfig: cm history} illustrates that varying $c_1/\Tilde{c}_1$ influences both the magnitude of the numerical pressure error and the amplitude of the temporal variations in the error. In general, larger values of $c_1/\Tilde{c}_1$ reduce the variations, as the contribution of the divergence of $\vb{u}^{(0)}$ to the pseudo-pressure is diminished.  At the same time, the (absolute) error values seem to slightly increase.
To quantify these trends, we calculate the absolute mean pressure error $| \Bar{\epsilon}_p |$ and root-mean-square (r.m.s.) pressure error $\epsilon_p'$ for different $c_1/\Tilde{c}_1$ values over the time interval $t = 0.7$ to $t = 1$, where the error is relatively steady. The results, shown in Figure~\ref{subfig: cm analysis}, indicate that $\epsilon_p'$ decreases by a factor of five between $c_1/\Tilde{c}_1 = 0$ (which corresponds to the algorithm used for stationary boundaries) and $c_1/\Tilde{c}_1 = 1$ (which completely removes the term $\nabla \cdot \mathbf{u}^{(0)}$ from the right-hand side of the pseudo-pressure Poisson's equation). Simultaneously the mean pressure error $| \Bar{\epsilon}_p |$ increases, though only by a factor of $\sim 1.8$ over the same range. Since the primary goal of our first-stage modification is to reduce pressure noise, we choose $c_1/\Tilde{c}_1 = 2/3$ as a compromise for the moving boundary simulations in the remainder of this work. As discussed in section~\ref{subsec: NS moving bc}, $c_i / \Tilde{c_i} = 0$ for the subsequent stages $i>1$. 

Moving on to the convergence tests of the moving star case, Figures~\ref{subfig: rotating star CFL u},~\ref{subfig: rotating star CFL p} and~\ref{subfig: rotating star CFL div} plot the $L_2$ and $L_{\infty}$ norms of the error fields at $t = 1$ for $u_x$, $p$ and $\vartheta$, respectively. The convergence results of the fourth order scheme $(5,4)$, and the lower order schemes $(3,4)$ and $(3,2)$, confirm that the conclusions from the static case remain valid for the moving boundary algorithm. Figure~\ref{subfig: rotating star long-time divergence} illustrates the long-time evolution of the $L_\infty$ norm of $\vartheta$ for the rotating star case. Due to the inherent noise in the raw data, as discussed above, and the fact that we evaluate the $L_{\infty}$ norm within a temporally changing domain, the data is smoothed for improved visualization. The remaining high frequency oscillations match the rotating period $T = 2$. The figure indicates that divergence remains stable over time for moving boundary simulations. 
Lastly, again we test the accuracy of pressure distribution using the same method as in the static case, with results shown in Figure~\ref{subfig: rotating star CFL pdis}. The results confirm that the pressure distribution on moving boundaries retains same accuracy as the pressure field itself. 

Overall, our convergence results in this section shows that the $(5,4)$ scheme is an effective and stable scheme for fourth-order simulations of the incompressible Navier-Stokes equations with stationary and moving immersed boundaries. In the next section, we will apply this algorithm to various flow scenarios, and compare its accuracy to the second-order scheme $(3,2)$. 

\section{Results part 2 : comparisons and conjugate heat transfer}
\label{sec: resultsP2}
In this section, we apply our method to a series of test cases for which exact solutions do not exist. The cases include both static and moving boundary simulations for which results are documented in literature, as well as a multi-physics application for Rayleigh-Bénard convection with immersed boundaries and conjugate heat transfer. We conduct long-time accuracy evaluations, a self-convergence analysis, and surface distribution studies to confirm the accuracy of the proposed method across the considered benchmark cases. Additionally, we compare the fourth-order $(5,4)$ scheme with the second-order $(3,2)$ scheme to discuss the advantages of high order immersed boundary treatment in Navier-Stokes algorithms.

\subsection{Flow past cylinder}
\label{subsec: flow past cylinder}
The flow past a static cylinder is a classic example for testing incompressible flow solvers. In our simulations, we consider a computational domain of size $[0, 2L) \times [0, L)$ for steady cases, and $[0, 3L) \times [0, L)$ for unsteady cases. We apply slip boundary conditions on the top and bottom surfaces of the domain, and inflow/outflow on the left/right sides of the domain. 
The cylinder has diameter $D = 0.0625L$ and is embedded and centered at $\vb{x}_c = (0.651L, 0.503L)$. 

\begin{figure}
    \begin{subfigure}{0.45\linewidth}
        \centering
        \includegraphics[width=1.0\textwidth]{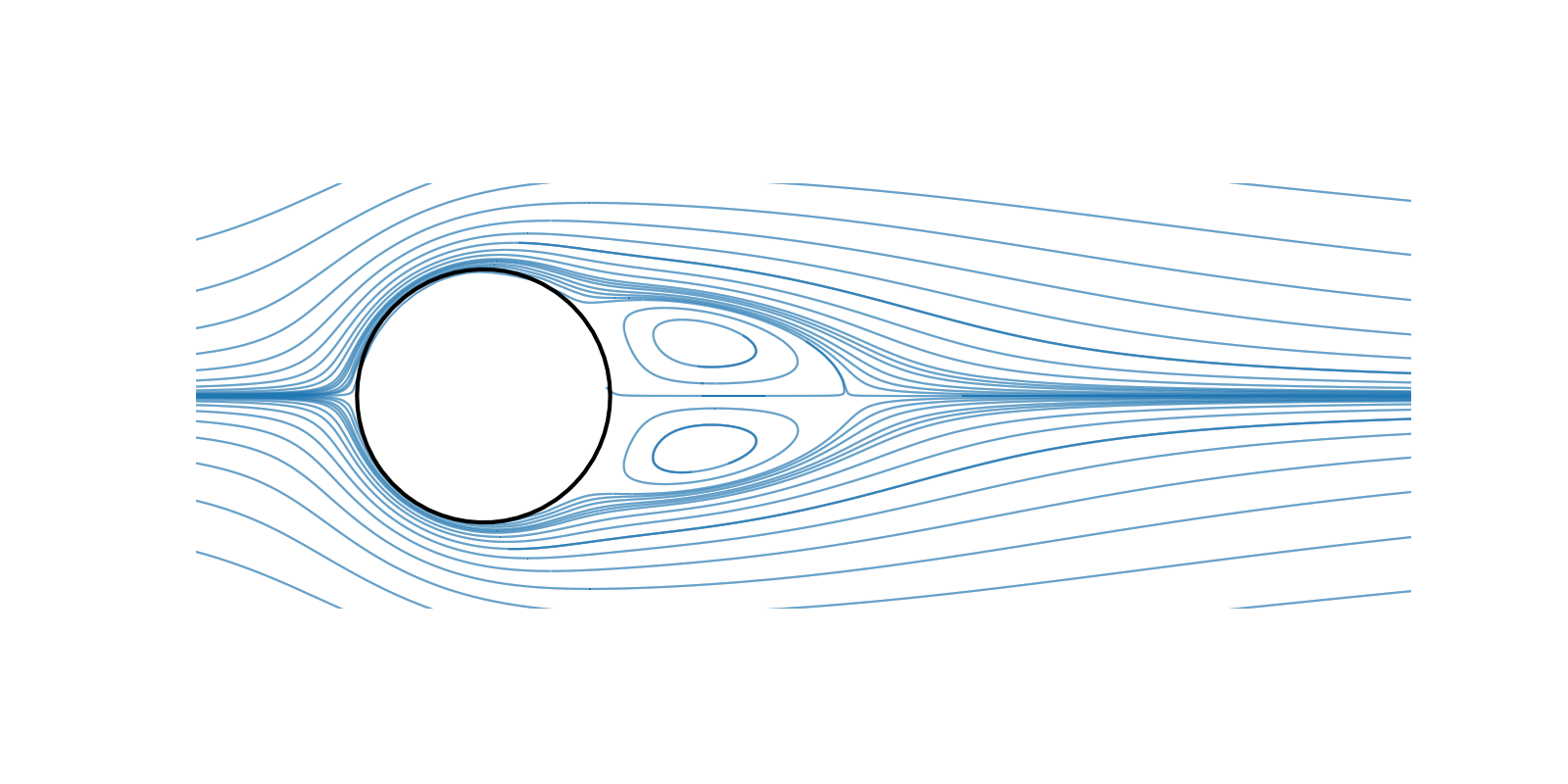}
        \caption{$Re = 20$.}
    \end{subfigure}
    \hspace{0.02\textwidth}
    \begin{subfigure}{0.45\linewidth}
        \centering
       \includegraphics[width=1.0\textwidth]{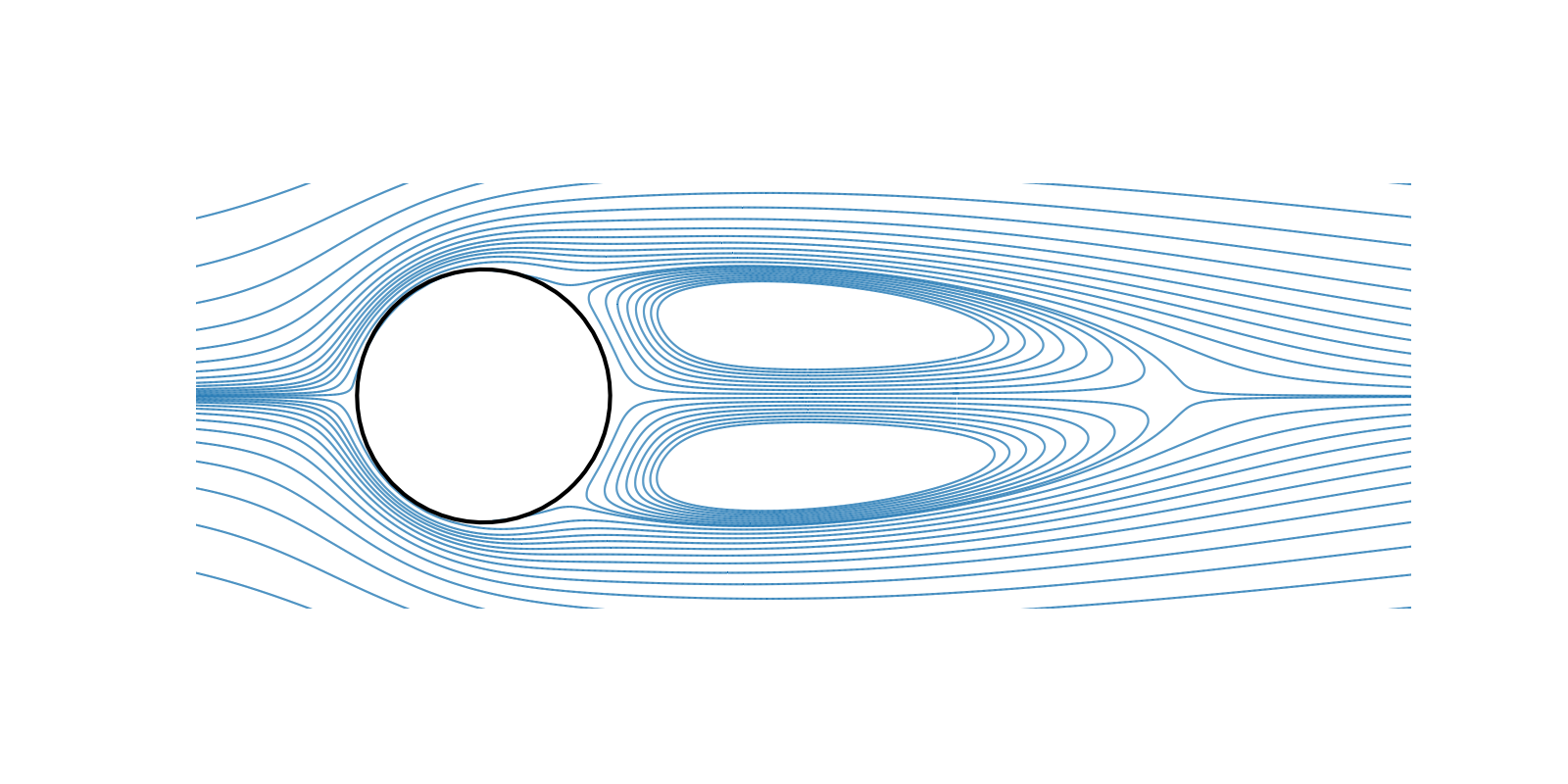}
        \caption{$Re = 40$.}
    \end{subfigure}

    \caption{Streamlines for the flow past a stationary cylinder at $\Tilde{t} = 90$, computed using scheme $(5,4)$.} 
    \label{fig: rotatingCylinder_streamlines}
\end{figure}

\begin{figure}

    \begin{subfigure}{0.43\linewidth}
        \centering
        \includegraphics[width=\textwidth]{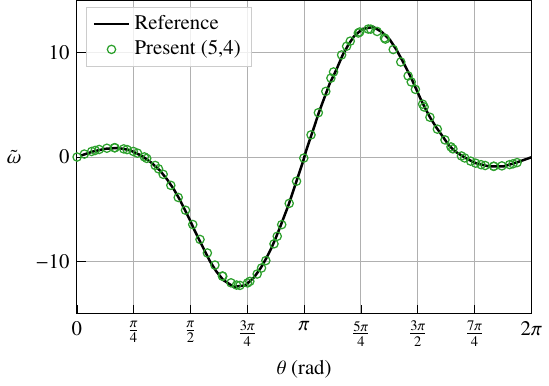}
        \caption{Vorticity distribution.}
    \end{subfigure}
    \hspace{0.02\textwidth}
    \begin{subfigure}{0.43\linewidth}
        \centering
        \includegraphics[width=\textwidth]{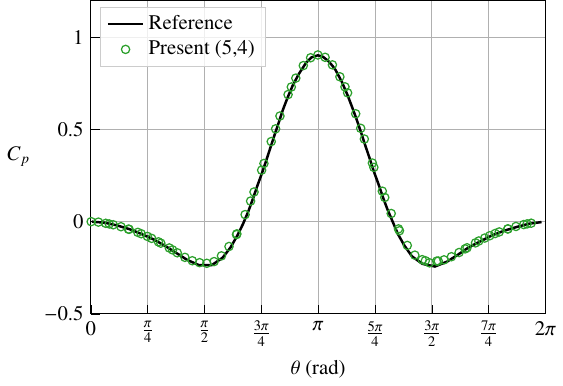}
        \caption{Pressure distribution.}
    \end{subfigure}

    \caption{Steady-state surface distribution of vorticity and pressure on the cylinder at $Re=40$. Fourth-order results using scheme $(5,4)$ are compared with reference results from \citet{Xu2006}.} 
    \label{fig: flowPastCylinder_steady}
\end{figure}

The equations are discretized using the $(5,4)$ algorithm, with the number of grid points in the domain set as $640 \times 320$ and $960 \times 320$, respectively, so that the non-dimensional spacing $h/D = 0.05$. For time integration we use a constant $CFL = |\vb{u}|_{max}\Delta t / h = 0.2$, where $|\vb{u}|_{max}$ is the maximum velocity magnitude in the computational domain.
We evaluate the flow at Reynolds numbers $Re = U_{\infty}D/\nu$ between $20$ and $200$. For the steady flows at $Re = 20$ and $Re = 40$, Figure~\ref{fig: rotatingCylinder_streamlines} shows the streamlines around the cylinder at $\Tilde{t} = U_{\infty} t/D = 90$. We compare the trailing bubble length $\Tilde{L}_{TB} = L_{TB}/D$, the separation angle $\theta_s$, and the drag coefficient $C_d = 2F_x/(\rho U_{\infty}^2 D)$ with previously reported values in literature in Table~\ref{tab: steadyFlowPastCylinder}. Here, $\theta_s$ is the angle measured counterclockwise from the leading stagnation point of the cylinder to the separation point with zero vorticity, while $L_{TB}$ represents the distance from the downstream stagnation point of the cylinder to the stagnation point in the wake where $u_x = 0$. Our results agree with reference solutions. Additionally, we compare the vorticity and pressure distributions along the cylinder surface at $Re = 40$ with the data from \citet{Xu2006} in Figure~\ref{fig: flowPastCylinder_steady}. The vorticity is non-dimensionalized as $\Tilde{\omega} = \omega D/U_{\infty}$,  and the pressure coefficient is given by $C_p = 2(p - p_0)/(\rho U_{\infty}^2)$, where $p_0$ is the pressure at $\theta = 0$. The surface distributions match the reference curves, demonstrating the ability of our sharp immersed method to accurately resolve surface quantities on immersed boundaries.

\begin{table}[htb]
    \centering
    \begin{tabular}{ | c | c  c  c | c  c  c |} 
    \hline
    Author  &  &$Re=20$ & & &$Re = 40$ & \\
    
            & $\Tilde{L}_{TB}$ & $\theta_s$ & $C_d$ & $\Tilde{L}_{TB}$ & $\theta_s$ & $C_d$ \\ 

    \hline

    Present $(5,4)$ & 0.93 & 43.7\textdegree & 2.23 & 2.23 & 53.8\textdegree & 1.67\\

    \citet{Calhoun2002}& 0.91 & 45.5\textdegree & 2.19 & 2.18 & 54.2\textdegree & 1.62\\

    \citet{Russell2003}& 0.94 & 43.3\textdegree & 2.13 & 2.29 & 53.1\textdegree & 1.60\\
    
    \citet{Xu2006}& 0.92 & 44.2\textdegree & 2.23 & 2.21 & 53.5\textdegree & 1.66\\
    \hline
    \end{tabular}
    \caption{Steady-state trailing bubble length $\Tilde{L}_{TB}$, separation angle $\theta_s$, and drag coefficient $C_d$ of the flow past cylinder case with $Re = 20$ and $Re = 40$, compared with previous computational results.}
    \label{tab: steadyFlowPastCylinder}
\end{table}

For simulations at $Re = 100$ and $Re = 200$ the flow is unstable. We evolve these simulations until $\Tilde{t} = 180$ and show the long-time statistics of the drag coefficient $C_d$ and the lift coefficient $C_l = 2F_y/(\rho U_{\infty}^2D)$ in Table~\ref{tab: unsteadyFlowPastCylinder}. Our results are compared with reference data, showing that all computed values fall within the reported ranges from previous studies.%
\begin{table}[htb]
    \centering
    \begin{tabular}{ | c | c  c  c | c  c  c |} 
    \hline
    Author  &  &$Re=100$ & & &$Re = 200$ & \\
    
            & $C_d$ & $C_l$ & $S_t$ & $C_d$ & $C_l$ & $S_t$ \\ 

    \hline

    Present $(5,4)$ & 1.42 $\pm$ 0.008& $\pm$0.31 & 0.175 & 1.21 $\pm$ 0.036 & $\pm$0.57 & 0.199\\

    \citet{Calhoun2002}& 1.33 $\pm$ 0.014 & $\pm$0.298  & 0.175 & 1.17 $\pm$ 0.058 & $\pm$ 0.67 & 0.202\\

    \citet{Russell2003}& 1.38 $\pm$ 0.007 & $\pm$0.300 & 0.169 & 1.29 $\pm$ 0.022 & $\pm$ 0.50 & 0.195\\

    \citet{Xu2006}& 1.423 $\pm$ 0.013 & $\pm$0.34 & 0.171 & 1.42 $\pm$ 0.04 & $\pm$ 0.66 & 0.202\\
    \hline
    \end{tabular}
    \caption{Long-time statistics of the drag coefficient, lift coefficient and Strouhal number of the flow past cylinder case with $Re = 100$ and $Re = 200$, compared with previous computational results.}
    \label{tab: unsteadyFlowPastCylinder}
\end{table}

\subsection{Pitching plate}
\begin{figure}
    \centering
    \begin{subfigure}[t]{0.3\linewidth}
        \centering
        \raisebox{25pt}{\includegraphics[width=1.0\textwidth]{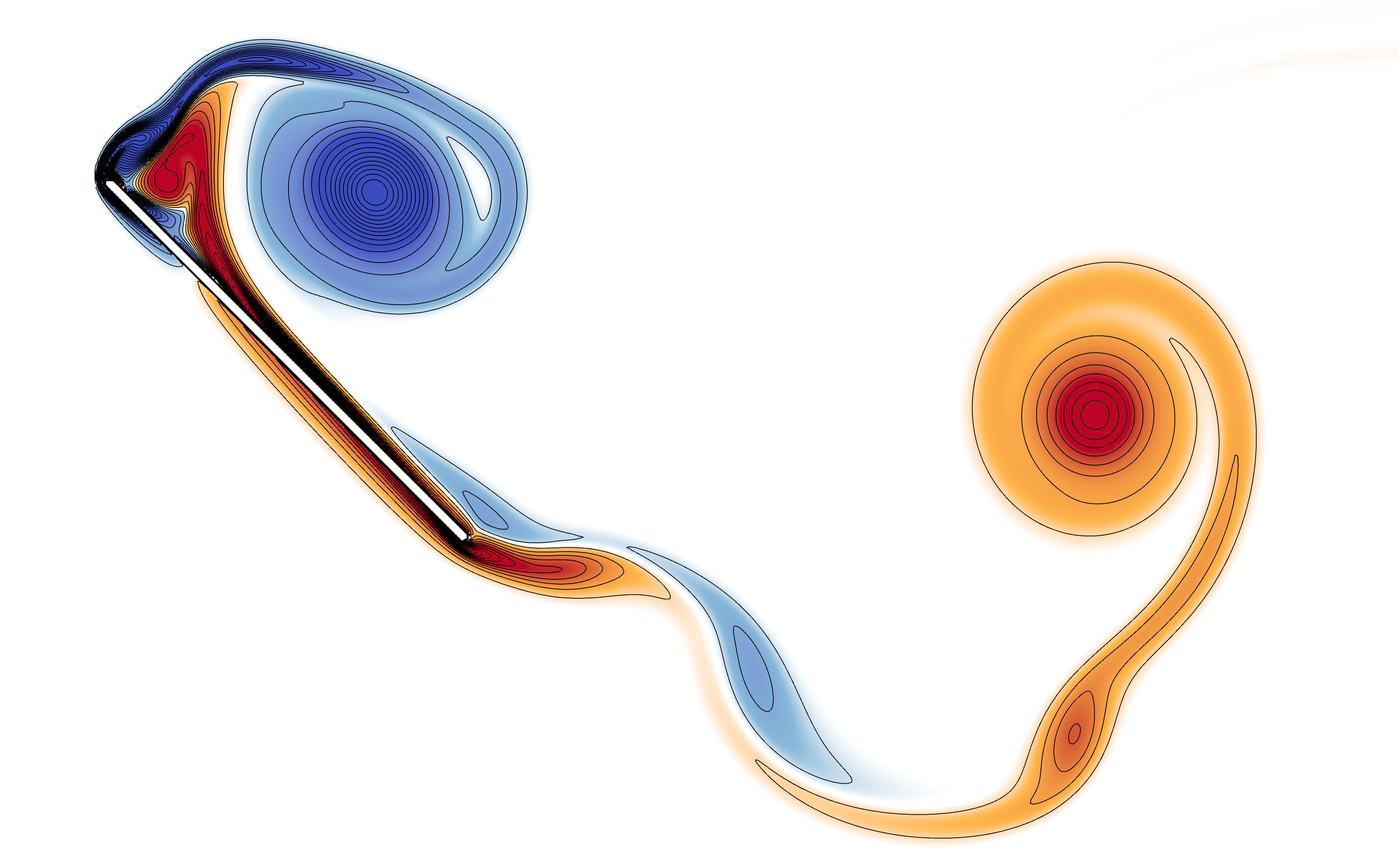}}
        \vspace{-25pt}
        \caption{Vorticity field  at $\Tilde{t} = 0.84$ for the $K=0.6$ case, simulated using the $(5,4)$ scheme with $\Tilde{N} = 192$.}
        \label{subfig: pitch contour}
    \end{subfigure}        
    \begin{subfigure}{0.32\linewidth}
        \centering
        \includegraphics[width=\textwidth]{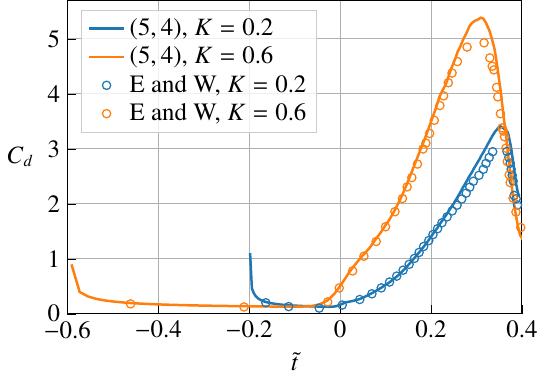}
        \caption{$C_d$ time evolution.}
    \end{subfigure}
    \hspace{0.02\textwidth}
    \begin{subfigure}{0.32\linewidth}
        \centering
        \includegraphics[width=\textwidth]{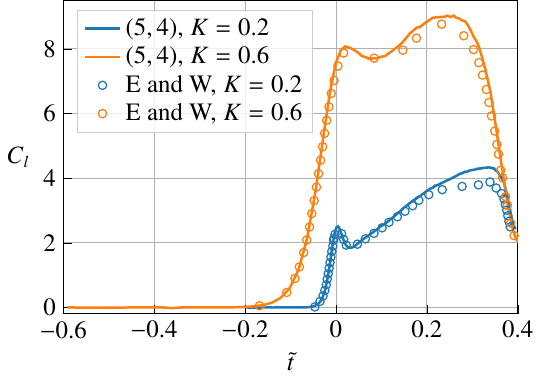}
        \caption{$C_l$ time evolution.}
    \end{subfigure}
    \caption{(a) Vorticity field for the pitching plate example at $Re=1000$. (b), (c) Evolution of $C_d$ and $C_l$ on the pitching plate with $Re = 1000$, as a function of non-dimensional time $\tilde{t}$. Results using our fourth-order $(5,4)$ scheme with resolution $c/h = 256$ are compared with reference results from~\cite{eldredge_high_fidelity_2010} (E and W).} 
    \label{fig: pitching reference}
\end{figure}

To validate our algorithm on an application with moving boundaries, we consider a pitching plate in a free-stream as analyzed in~\cite{eldredge_high_fidelity_2010}. For this case, the computational domain is of size $[0, 1.25L) \times [0, L)$ with resolution $1.25N \times N$. The plate has a chord length $c = 0.25L$ and a thickness $0.023c$, with semicircular edges. The Reynolds number of the flow is $Re = U_{\infty}c/\nu = 1000$. After an initial transient the plate rotates around its leading edge at $\vb{x}_b = (1.73c, 2.21c)$ with an angle of attack $\alpha(t)$ that is ramped up from zero following:

\begin{equation}
    \alpha(t) = \alpha_0 \frac{G(t)}{G_{max}},
    \label{eq: rotating angle 1}
\end{equation}
where $\alpha_0$ is set as 45 degrees. The function $G(t)$ is the smooth ramp-up and ramp-down function from~\cite{eldredge_high_fidelity_2010}:

\begin{equation}
    G(t) = \ln\left[{\frac{\cosh(a_sU(t - t_1)/c) \cosh(a_s U(t - t_4)/c)}{\cosh(a_s U(t - t_2)/c)\cosh(a_s U (t - t_3)/c)}}\right], \quad G_{max} = 2 a_sU(t_2 - t_1)/c,
    \label{eq: rotating angle 2}
\end{equation}
where $a_s$, controlling the speed of the kinematic transitions, is set to 11 for all simulations. The times $t_1$, $t_2$, $t_3$, and $t_4$ mark the transition stages of the rotation. Here $t_1 = c/U_{\infty}$ and $t_2 = t_1 + \alpha_0/\dot{\alpha}_0$, with the rotation rate expressed non-dimensionally through $K = 0.5\dot{\alpha}_0 c/U_{\infty}$. The ramp-down times are $t_3 = t_2 + 1.12c/U_{\infty}$, and $t_4 = t_3 + (t_2 - t_1)$; however, we stop our simulations at $t = 2c/U_{\infty}$ so that $t_3$ is never reached. We simulate cases with non-dimensional rotation rates $K = 0.2$ and $K = 0.6$, and express results using dimensionless time $\Tilde{t} = K(tU_{\infty}/c -1)$; an example of the late-time flow field at $K=0.6$ is shown in Figure~\ref{subfig: pitch contour}.

For each case, we evaluate the drag coefficient $C_d = 2F_x/(\rho_f U_{\infty}^2c)$ and lift coefficient $C_l = 2F_y/(\rho_f U_{\infty}^2c)$. Figures~\ref{fig: pitching reference}b and  ~\ref{fig: pitching reference}c compare the $C_d$ and $C_l$ history computed using  the $(5,4)$ scheme under a resolution of $\Tilde{N} = c/h = 256$ with the results in~\cite{eldredge_high_fidelity_2010}. Our results show good qualitative agreement with the reference data. A more quantitative comparison with \cite{eldredge_high_fidelity_2010}, or with our own higher resolution results in \cite{ji_sharp_2023}, is not useful as both these reference results rely on vorticity-velocity formulations with free-space domain boundary conditions. In contrast, our approach uses the inflow/outflow conditions~\eqref{eq:farfield_bc}, so that the numerical results are expected to converge to a slightly different state. 

\begin{figure}
    \centering
    \begin{subfigure}[t]{0.32\linewidth}
        \centering
        \includegraphics[width=\textwidth]{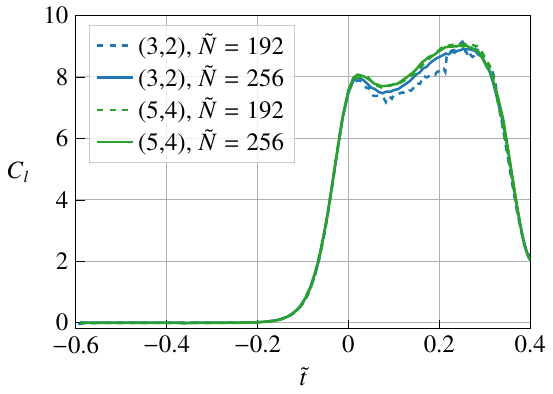}
        \caption{Time evolution of $C_l$.}
        \label{subfig: drag comparison}
    \end{subfigure}
    \hspace{0.02\textwidth}
    \begin{subfigure}[t]{0.32\linewidth}
        \centering
        \includegraphics[width=\textwidth]{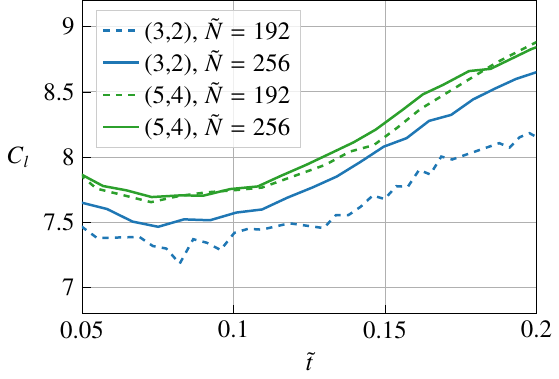}
        \caption{Zoom in of plot (a) for $0.05 \le \tilde{t} \le 0.2$.}
    \end{subfigure}
    \hspace{0.02\textwidth}
    \begin{subfigure}[t]{0.3\linewidth}
        \centering
        \includegraphics[width=\textwidth]{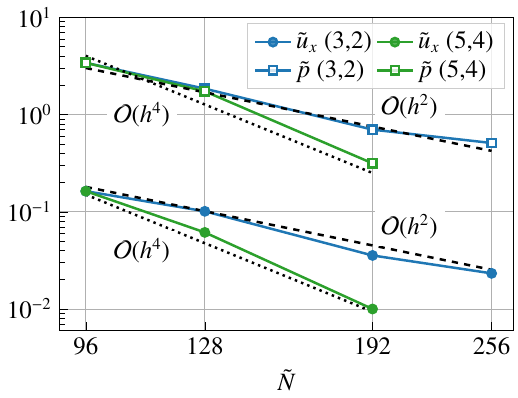}
        \caption{Self-convergence plot of the $L_{\infty}$ norm of the error in the $x$-velocity component and pressure fields, compared to the $(5,4)$ solution at $\tilde{N} = 256$.}
        \label{subfig:pitching_self_convergence_Linf}        
    \end{subfigure}    
    \caption{Comparison of results from the $(5,4)$ scheme and the $(3,2)$ scheme across resolutions $\Tilde{N}$, for the pitching plate case with $K = 0.6$.} 
    \label{fig: pitching comparison}
\end{figure}

Nevertheless, we can use our own results to evaluate the effect of discretization order. To do so, we simulate the $K = 0.6$ case using both $(3,2)$ and $(5,4)$ schemes separately, across varying resolutions. The drag histories for different configurations are shown in Figure~\ref{subfig: drag comparison}. The results indicate that all cases converge to the highest resolution solution obtained by the $(5,4)$ scheme at $\Tilde{N} = 256$. In fact, the $(5,4)$ scheme is already converged at $\Tilde{N} = 192$, whereas the $(3,2)$ scheme still exhibits noticeable variations even at $\Tilde{N} = 256$. A linear reduction in resolution such as $(192/256)$, for 2D simulations at constant CFL, implies a $(256/192)^3 \sim 2.4$ times reduction in the corresponding spatio-temporal degrees of freedom. Figure~\ref{subfig: drag comparison} implies that the second-order scheme requires at least a resolution of $c/h=384$ for convergence, so we can estimate that the fourth order scheme reduces the number of spatio-temporal degrees-of-freedom required for a converged solution by roughly an order of magnitude. Of course, as the desired numerical error decreases, this gap will continue to grow. Moreover, in 3D, the cubic power would be raised to a quartic one, further improving the computational efficiency of the fourth-order scheme.

Finally, Figure~\ref{subfig:pitching_self_convergence_Linf} shows a self-convergence plot of both the second order $(3,2)$ scheme, and the fourth-order $(5,4)$ scheme, compared to the results of the $(5,4)$ scheme at $\Tilde{N}=256$. We report non-dimensional errors in the $x$-velocity $\Tilde{u}_x = u_x/U_{\infty}$, and the pressure $\Tilde{p} = (p - \Bar{p})/(\rho U_{\infty}^2)$, where $\Bar{p}$ is the mean value of the pressure field. The errors are measured at $\Tilde{t} = 0$ and we report the $L_\infty$ error norm. The results show that both the velocity field and the pressure field self-converge at the expected orders everywhere in the domain.

\subsection{Vortex dipole impinging on immersed wall}
\label{sec:vortexdipole}
The vortex dipole-wall collision has been a benchmark problem for 2D Navier-Stokes solvers starting with the work of \cite{Clercx:2006a}. It is known as a challenging test case to converge, because the late stage evolution of the dipole strongly depends on the thin boundary layer of vorticity generated at the wall. To the best of our knowledge, the only results using an immersed method were presented in \cite{Keetels:2007} using a  penalization method. The authors showed converged results at $Re=1000$ using a $2730^2$ resolution grid, with the immersed walls aligned with the cartesian coordinate directions.

Here, we evaluate the dipole-wall collision at $Re=1000$ and $Re=1250$ using our immersed method. We define the flow domain using an immersed, rounded square rotated at an angle of $3\pi/10$ with respect to the Cartesian grid. The square has length $L = 0.65123$ and is placed within a unit square computational domain, centered on the point $(0.501,0.499)$. The size, center, and rotation angle are chosen to ensure the immersed domain boundaries intersect the background grid non-trivially across the domain. Further, the immersed domain has rounded corners with a radius of curvature of $L/10$ to prevent any issues with the immersed discretization. This radius of curvature is much larger than what the IIM requires to comply with~Equation~\eqref{eq:curvature_condition}, but since the reference results show relative insensitivity to the lateral domain boundary conditions, this should not cause issues. For comparison, we use the vorticity contours at $Re=1000$ from the spectral reference method in \cite{Keetels:2007}. Further, we repeat the same simulation at $Re=1250$ to be able to compare with the wall vorticity distribution reported in \cite{Clercx:2006a}. 

The dipole is initialized according to the initial conditions described in \cite{Clercx:2006a,Keetels:2007}. We simulate the flow with the fourth-order $(5,4)$ and the second-order $(3,2)$ schemes using a CFL number of $0.4$ and resolutions of $1024^2$ and $1536^2$ for the entire unit square domain. The effective resolution within the immersed domain associated with these values is approximately $667^2$ and $1000^2$, respectively. For each simulation, we visualize the vorticity contours at $t=0.5$ (after the first collision) and $t=0.8$ (after the second collision), and rotate and scale them to account for our slanted, scaled-down immersed domain. The contours are then overlaid on the figures provided in \cite{Keetels:2007}. 

\begin{figure}
\includegraphics[width=\textwidth]{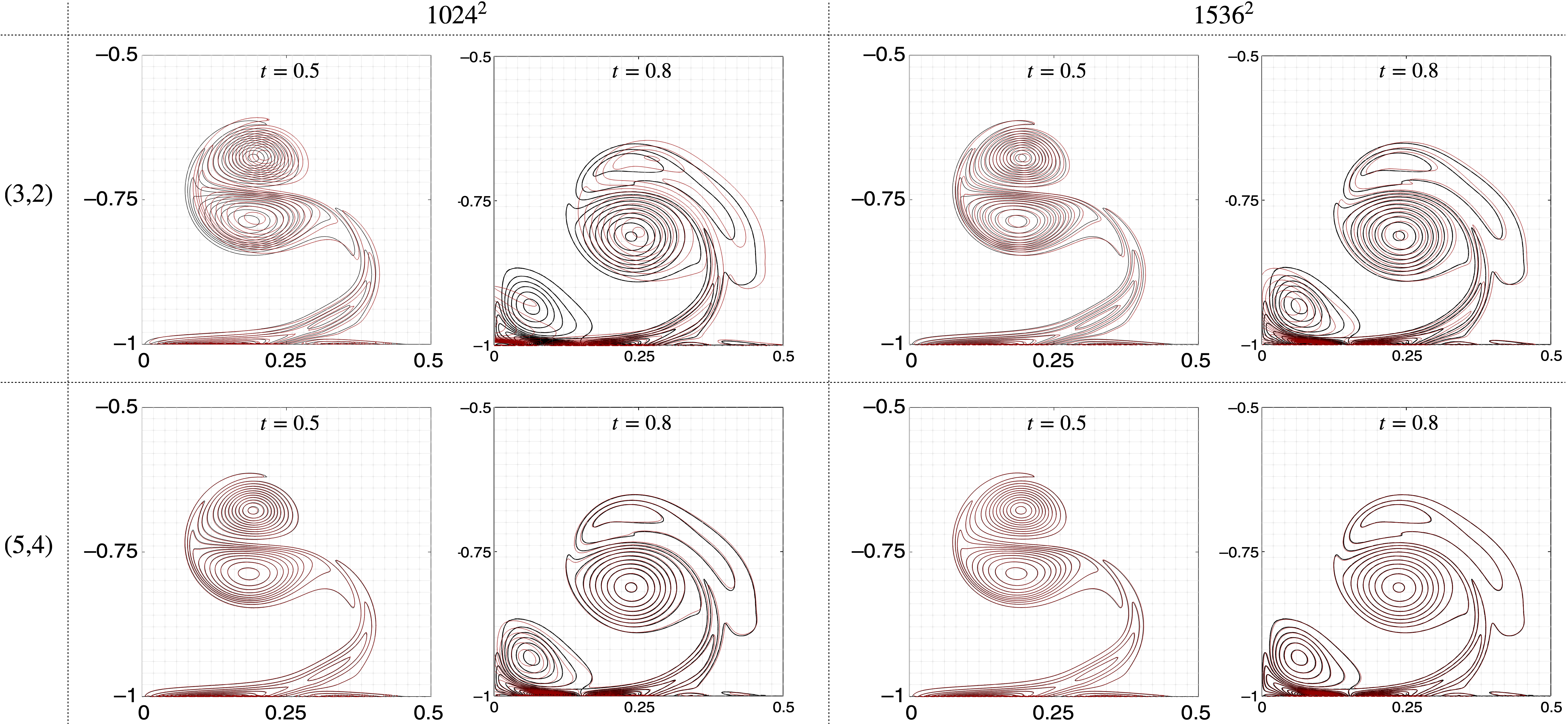}
    \caption{Comparison of vorticity contours of the dipole-wall collision at $Re=1000$ between the spectral reference \cite{Keetels:2007} (black) and the immersed method (red), at $t=0.5$ and $t=0.8$. The top row shows results from the second order $(3,2)$ formulation, the bottom row from the fourth order $(5,4)$ formulation. The left column is at resolution $1024^2$ (effectively $667^2$ in the immersed domain), the right column is at resolution $1536^2$ (effectively $1000^2$ in the immersed domain).} 
    \label{fig:dipole_contours}
\end{figure}

Figure~\ref{fig:dipole_contours} shows our simulated vorticity contours (in red) on top of the reference contours (in black). We see that the low order scheme (top) does not match the reference well at domain resolution $1024^2$, especially near the wall; at $1536^2$ the contours match better but are not yet converged. In contrast, the $(5,4)$ scheme provides very good agreement at $1024^2$ except near the wall at late time; at $1536^2$ the fourth-order results are indistinguishable from the spectral simulation at both time instances. This broadly is consistent with the results of the pitching plate example, where the second order scheme required 1.5--2 times the linear resolution of the fourth order scheme to converge to a similar accuracy.

\begin{figure}
\includegraphics[width=\textwidth]{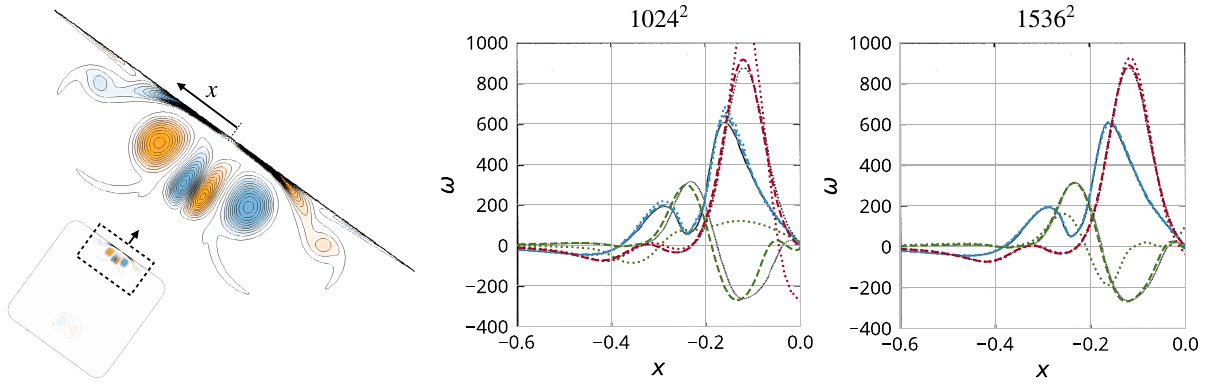}
    \caption{Left: vorticity contours at $t=0.6$ for the $1536^2$, $(5,4)$ simulation of the dipole-wall collision at $Re=1250$, and definition of the direction of the wall coordinate $x$. Right: comparison of the wall vorticity as a function of $x$ between the spectral reference \cite{Clercx:2006a} (black lines) and the immersed method (colored lines) at $t=0.4$ (blue), $t=0.6$ (red), and $t=1.0$ (green). The immersed method results are shown for the fourth order $(5,4)$ (dashed) and the second order $(3,2)$ (dotted) algorithms at domain resolution $1024^2$ (center, effectively $667^2$ in the immersed domain) and $1536^2$ (right, effectively $1000^2$ in the immersed domain).}
    \label{fig:dipole_wall_vorticity}
\end{figure}

To compare the wall vorticity, we extract the velocity gradients on the wall at all control points using least-squares stencils containing both the flow points, and the no-slip boundary condition at the wall. From the wall velocity gradients we compute the associated wall vorticity $\omega_w = \partial(\vb{u} \cdot \vb{t})/\partial n$ with $\vb{t}$ the tangent vector, which in 2D is proportional to the viscous traction and thus governs the shear force acting on the wall. We then plot this quantity as a function of the wall coordinate. The results from the $(5,4)$ and $(3,2)$ algorithms at domain resolutions $1024^2$ and $1536^2$ are plotted on top of the results from \cite{Clercx:2006a}, at times $t=0.4$, $t=0.6$, and $t=1.0$ in Figure~\ref{fig:dipole_wall_vorticity}. For the second order algorithm (dotted) results at $1024^2$ are significantly off; at $1536^2$ the early time results match better, but the late time vorticity at $t=1.0$ does not compare well. The fourth order results (dashed) at $1024^2$ resolution (effectively $667^2$ in the immersed domain) capture the overall trends well up to $t=0.6$, but is slightly off at peak values. At late time $t=1.0$, the differences are more significant, though much better than the second-order results. The fourth-order $1536^2$ resolution results match very well with the reference spectral results, even at late time. This demonstrates the ability of the fourth order immersed method to capture high fidelity velocity gradients on the immersed wall. 

\subsection{Conjugate heat transfer}
\label{sec: showcase}
In this final example, we extend our high-order incompressible Navier–Stokes solver to simulate buoyancy-driven conjugate heat transfer. Buoyancy effects are incorporated using the Boussinesq approximation, which introduces a temperature-dependent body force in the momentum equations. The temperature field evolves according to a simple advection-diffusion equation in the flow domain, and a pure diffusion equation in the solid domain; conjugate conditions are imposed on the fluid-solid interface. Considering the fluid domain $\Omega^{+}$ and the solid domain $\Omega^{-}$, we solve the following non-dimensional equations~\cite{liu_from_rayleigh_2020}:

\begin{equation}
\begin{aligned}
    \frac{\partial \Tilde{\vb{u}}}{\partial \Tilde{t}} + \Tilde{\vb{u}} \cdot \nabla \Tilde{\vb{u}} &= -\nabla \Tilde{p} + \sqrt{\frac{{Pr}^{+}}{{Ra}^{+}}} \Delta \Tilde{\vb{u}} + T \vb{e}_y, \quad  &\vb{x} \in \Omega^{+}, \\
    \nabla  \cdot \Tilde{\vb{u}} &= 0, \quad &\vb{x} \in \Omega^{+}, \\
    \Tilde{\vb{u}}  &= \vb{0}, \quad & \vb{x} \in \Gamma, \\
    \frac{\partial T}{\partial \Tilde{t}} + \nabla \cdot (\Tilde{\vb{u}}T) &= \frac{1}{\sqrt{{Pr}^{+} {Ra}^{+}}}\Delta T,  \quad &\vb{x} \in \Omega^{+}\\
    \frac{\partial T}{\partial \Tilde{t}} &= \frac{1}{\sqrt{{Pr}^{-} {Ra}^{-}}}\Delta T, \quad &\vb{x} \in \Omega^{-}, \\
    [T] &= 0, \quad & \vb{x} \in \Gamma, \\
    [\kappa \partial_n T] &= 0 \quad & \vb{x} \in \Gamma,
\end{aligned}
\label{eq:conjheat}
\end{equation}
where the thermal diffusivity is piecewise-continuous in the fluid and solid domains
\[
\kappa(\vb{x}) = \begin{cases} \kappa^+ & \vb{x} \in \Omega^+, \\ \kappa^- & \vb{x} \in \Omega^-.
\end{cases}
\]

These governing equations are non-dimensionalized with $L$ for length, $\delta$ for temperature, and $U = \sqrt{g\alpha_T \delta L}$ for velocity, where $g$ is gravitational acceleration and $\alpha_T$ is the isobaric expansion coefficient. Therefore, $\Tilde{\vb{u}} = \vb{u}/U$, $\Tilde{p} = p/(\rho U^2)$, $\Tilde{t} = tU/L$, and $\Tilde{\omega} = \omega L/U$. Further, $\vb{e}_y$ is the unit vector in the vertical direction. The two dimensionless parameters governing the fluid flow are the Rayleigh number, ${Ra}^{+} = g\alpha_T \delta L^3/(\nu \kappa^{+})$, and the Prandtl number ${Pr}^{+} = \nu/\kappa^{+}$, where $\kappa^{+}$ is the thermal diffusivity of the fluid. For the solid, the governing non-dimensional quantities are  ${Ra}^{-} = (\kappa^+/\kappa^{-}) Ra^{+}$ and ${Pr}^{-} = (\kappa^+/\kappa^-) Pr^{+}$.

 To solve this system we include the buoyancy effects at each stage $i$ by computing $T^{(i-1)}\vb{e}_y$ and adding it to the advection and diffusion terms in Equation~\eqref{eq: step-1}. The Runge-Kutta integrator then updates $T^{(i)}$ from $\vb{u}^{(i-1)}$ and $T^{(i-1)}$ using the same IIM-corrected differential operators for advection and diffusion as the Navier-Stokes system. The conjugate conditions on the immersed interface are enforced using our IIM method as explained in section~\ref{subsec:FD_IIM}. This solution method for the temperature field has been verified extensively in our previous work on the advection-diffusion equation with discontinuous coefficients \cite{gabbard2023high, gabbard_2024}, though a prescribed velocity field was used in those cases.
 
\begin{figure}
    \centering
    \begin{subfigure}[t]{0.28\linewidth}
        \centering
        \includegraphics[width=1\textwidth]{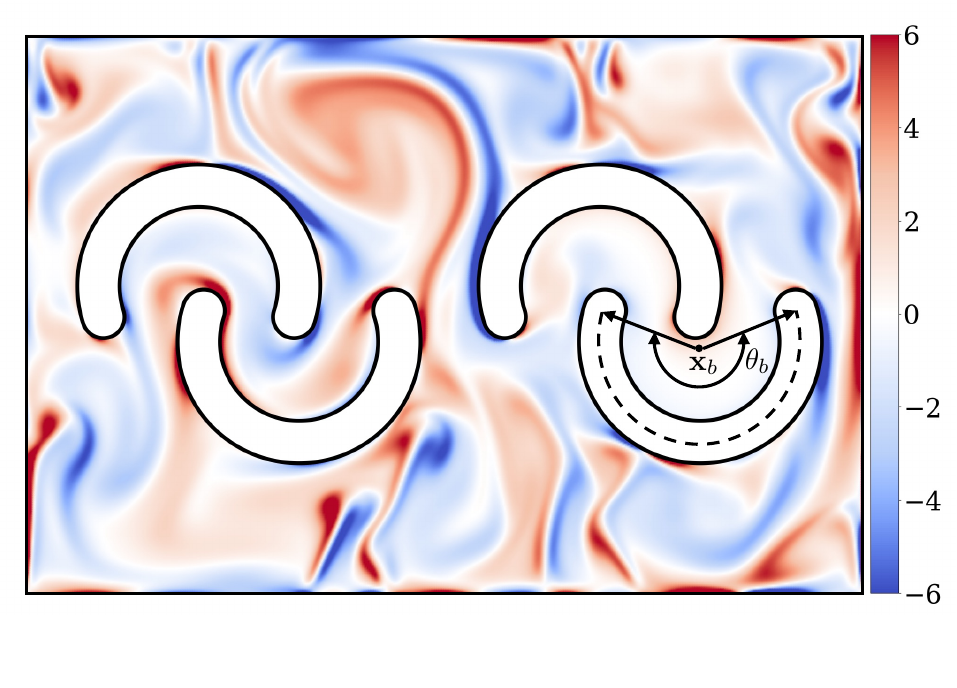}
        \caption{$\Tilde{\omega}$.}
        \label{subfig: heat_vort}
    \end{subfigure}
    \hspace{0.01\textwidth}
    \begin{subfigure}[t]{0.29\linewidth}
        \centering
        \includegraphics[width=1\textwidth]{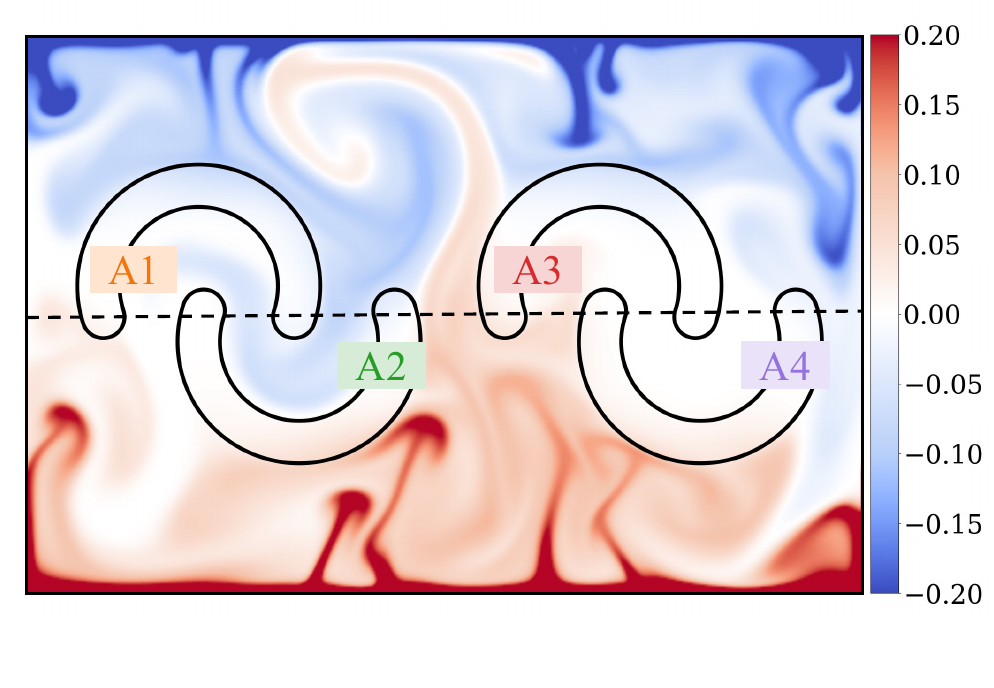}
        \caption[]{$T$ field and the centerline $y/L = 0.5$ (black dashed). }
        \label{subfig: heat_temp}
    \end{subfigure}
    \hspace{0.01\textwidth}
    \begin{subfigure}[t]{0.375\linewidth}
        \centering
        \includegraphics[width=\textwidth]{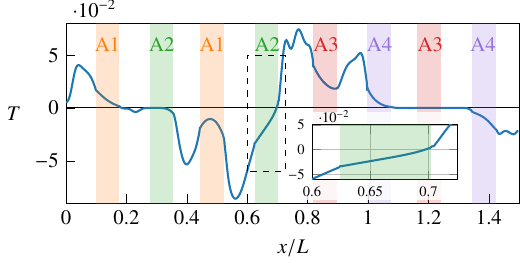}
        \caption{$T$ distribution along the centerline. Inset shows a zoom around the second A2 crossing as indicated by the dashed black square.}
        \label{subfig: heat across line}
    \end{subfigure}
    \caption{Results for the conjugate heat transfer simulation at $\Tilde{t} = 18$ with $N = 320$ using the $(5,4)$ scheme. The inset in (c) highlights the sharply resolved slope discontinuity of the temperature field at fluid-solid interfaces.} 
    \label{fig: heat fields}
\end{figure}

For this example, we consider a Rayleigh-Bénard problem, where a buoyancy-driven flow is induced by fluid layers heated from below and cooled from above. Due to the temperature gradient, an uneven density profile develops within the Rayleigh-Bénard cell, and the resulting buoyancy drives convective heat transport~\cite{ahlers_heat_2009}. Specifically, we simulate  a rectangular domain $[0, 1.5L) \times [0, L)$ discretized with a uniform resolution of $h = L/N$, embedding four static thickened arcs labeled A1 through A4. Each arc is defined by a center position $\vb{x}_b$, an inner radius $r_a = 0.142L$, an outer radius $r_b = 0.218L$, and an opening angle $\theta_b = 6\pi/5$, as illustrated in Figure~\ref{subfig: heat_vort}. Arcs A1 and A2 are located at $\vb{x}_b = (0.31L, 0.55L)$ and $\vb{x}_b = (0.49L, 0.45L)$, respectively, while arcs A3 and A4 are positioned $0.72L$ horizontally to the right of A1 and A2. For these simulations, we set ${Ra}^{+} = 10^8$ and ${Pr}^{+} = 2$, and use a diffusivity ratio of $\kappa^{+} / \kappa^{-} = 1/3$. All domain boundaries are considered as no-slip walls for the velocity field. For the temperature field, a homogeneous Neumann boundary condition is enforced on side walls with $\partial_n T = 0$. Dirichlet boundary conditions are applied to the temperature field at the top and bottom boundaries, with non-dimensional values $- 0.5$ and $ 0.5$, respectively. 

 The velocity field is initialized at rest. For the temperature field we introduce an initial perturbation near the walls to break symmetry:
\begin{align}
    T(\vb{x}, t=0) =
    \left \{ 
    \begin{array}{ll}
        0.5 - \dfrac{0.5 y}{0.01L\left( 2+ \sin{(8\pi x/L)}\right)},   &y < 0.01L\left( 2+ \sin{(8\pi x/L)}\right), \\[12pt] 
          -0.5 - \dfrac{0.5 (y - L)}{0.01L\left( 2+ \sin{(8\pi x/L)}\right)}, &y > L - 0.01L\left( 2+ \sin{(8\pi x/L)}\right), \\[12pt] 
         0, & \text{else.} 
    \end{array}
    \right .
\end{align}

The simulations are run with the fourth-order scheme $(5,4)$ and the second-order scheme $(3,2)$, using CFL = $0.2$. Figures~\ref{subfig: heat_vort} and~\ref{subfig: heat_temp}
 show the fourth-order vorticity and temperature fields at time $\Tilde{t} = 18$ with resolution $N = 320$. The Reynolds number at this time is $Re = |\Tilde{\vb{u}}_{max}| \sqrt{Ra/Pr} = 2035$, where $\Tilde{\vb{u}}_{max}$ denotes the maximum velocity magnitude in the domain. In Figure~\ref{subfig: heat across line} we show the temperature distribution along the centerline of the domain ($y/L = 0.5$) at this instance from the simulation. This centerline coincides with a grid line, so we include temperature values at the control points $\vb{x}_c$ in the plot. The plot shows the discontinuous slopes in the temperature field across fluid-solid interfaces. Consequently, this approach provides a high-order solution of complex multiphysics problems with interface-based discontinuities. 

\begin{figure}
    \begin{subfigure}[t]{0.38\linewidth}
        \centering
        \includegraphics[width=\textwidth]{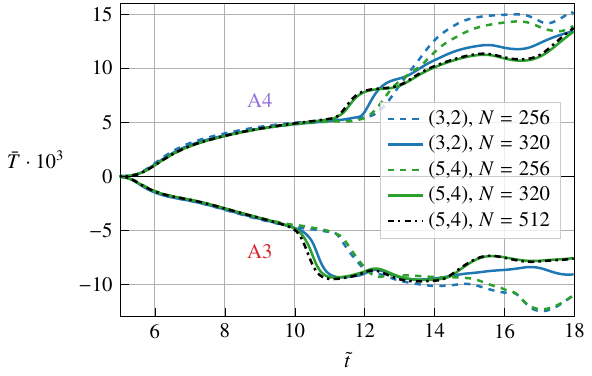}
        \caption{Time evolution of the mean temperature $\Bar{T}$ on the surface of A3 (bottom) and A4 (top) at different resolutions $N$ and schemes. }
        \label{subfig: heat convergence}
    \end{subfigure}
    \hspace{0.01\textwidth}
    \begin{subfigure}[t]{0.29\linewidth}
        \centering
        \includegraphics[width=1\textwidth]{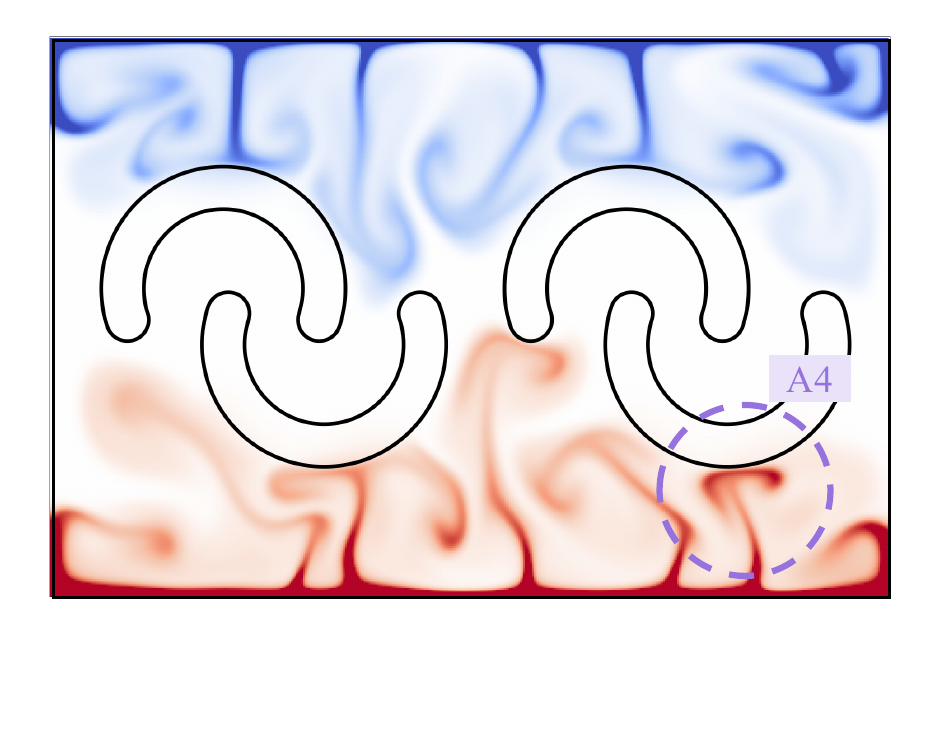}
        \caption{$T$ field at $\Tilde{t} = 12$ ($N = 320$ , $(3,2)$ algorithm).}
        \label{subfig: heat snapshot 32}
        \end{subfigure}
    \hspace{0.01\textwidth}
    \begin{subfigure}[t]{0.29\linewidth}
        \centering
        \includegraphics[width=1\textwidth]{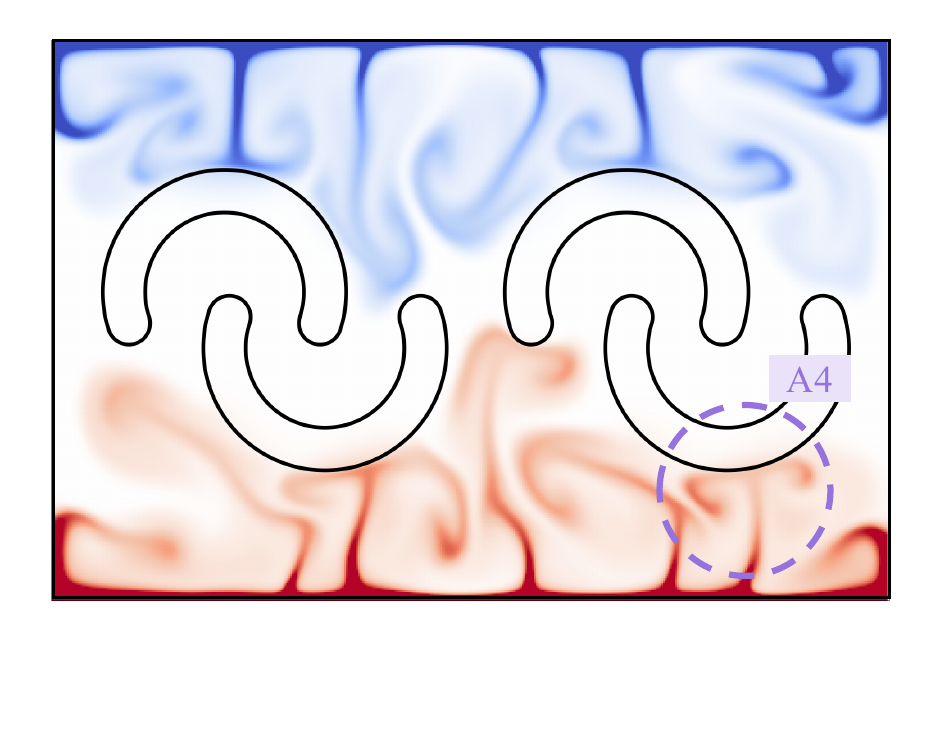}
        \caption{$T$ field at $\Tilde{t} = 12$ ($N = 320$ , $(5,4)$ algorithm).}
        \label{subfig: heat snapshot 54}
    \end{subfigure}
    
    \caption{The mean temperature evolution on arcs A3 and A4 (Panel a) show differences between the second and fourth order results starting around $\Tilde{t}=12$. Panels (b) and (c) show the physical variations in the temperature field explaining the temperature deviations on arch $A4$.}
    \label{fig: heat distribution}
\end{figure}

To evaluate the influence of high-order schemes on long-time simulations, we track the average temperature $\Bar{T}$ at the surfaces of A3 and A4 throughout the simulation. We compare results obtained using the $(5,4)$ and $(3,2)$ schemes at resolutions $N = 256$ and $N = 320$ against a reference simulation at $N = 512$ with the $(5,4)$ scheme, as shown in Figure~\ref{subfig: heat convergence}. The results indicate that the high-order $(5,4)$ scheme achieves convergence by $\Tilde{t} = 18$ at a resolution of $N = 320$, while under the same conditions, the $(3,2)$ scheme begins to deviate from the converged solution at $\Tilde{t} = 12$ for arc A4. Figures~\ref{subfig: heat snapshot 32} and~\ref{subfig: heat snapshot 54} show snapshots of the temperature field at $\Tilde{t} = 12$ using the $(3,2)$ and $(5,4)$ schemes, respectively. Comparing the two, we observe that in the $(5,4)$ scheme, a high temperature plume reaches arc A4 earlier than in the $(3,2)$ scheme, leading to qualitative difference in the late-time results. 
These results demonstrate that the fourth order method achieves convergence at lower resolutions in long-time simulations of highly nonlinear systems, making it particularly effective for accurate long-time simulations.

\section{Conclusion}
\label{sec: conclusion}
In this work, we present and analyze a fourth-order incompressible velocity-pressure Navier-Stokes solver using the immersed interface method (IIM), which can accurately simulate flows with static and moving boundaries as well as conjugate heat transfer. To achieve this, we propose a novel fifth-order IIM advection discretization scheme and a new high-order Runge-Kutta-based fractional step method for steady and unsteady boundaries. The solver is further extended to incorporate buoyancy-driven conjugate heat transfer. Extensive numerical experiments confirm up to fourth order accuracy for all variables under fixed-CFL conditions.

Through comparisons with lower order schemes, we quantify the effects of the high-order discretization on the quality of the solution. The results demonstrate that in our test cases, linear resolutions required for practical convergence using the fourth-order scheme are typically half to two-thirds of those for our second-order scheme. In 2D with fixed CFL, this translates to a four-to-eight reduction in spatio-temporal degrees-of-freedom required, thus significantly improving computational efficiency. In 3D, this should increase to 5--16 times fewer degrees-of-freedom. Naturally, as the desired accuracy increases, these improvements will grow further. For long-time simulations, the use of high-order far-field schemes in combination with a low-order immersed method was already shown to be beneficial in \cite{laizet_high-order_2009}; those same benefits will translate to our fully high order scheme as well, while further providing high-fidelity on- and near the immersed geometries. On the other hand, the computational cost of the fourth-order scheme naturally increases as it requires an extra Poisson equation per time step (due to the higher-order Runge-Kutta scheme) and more floating point operations; however these increases are far outweighed by the increased efficiency.

In addition to these contributions, the proposed algorithm retains several advantages from our high-order IIM method that have already been demonstrated~\cite{gabbard2023high, gabbard_2024}. First, the elliptical IIM scheme allows for the simulations of both convex and concave boundary geometries, providing geometric flexibility. Second, beyond Dirichlet and Neumann boundary conditions, the IIM method is able to enforce jump boundary conditions, enabling the extension of the solver to high-order multiphysics problems, as demonstrated in our conjugate heat transfer examples. More general boundary conditions could easily be implemented to impose wall or interface stress conditions \cite{Gabbard2025b}, as required e.g.\ in wall-models of Large-Eddy Simulations, multiphase flows, or fluid-structure interactions. Third, the IIM is easily integrated into high order adaptive grid refinement frameworks, such as the one proposed in \cite{Gillis:2022}; in fact, our earlier work already demonstrated high order 3D adaptive-grid solutions to advection-diffusion equations with moving immersed boundaries \cite{gabbard_2024}. Overall, this work significantly expands the ability of immersed methods to efficiently yield high-fidelity solutions for incompressible-flow-based multiphysics problems, while also providing a extensible foundation for more complex problems.

\section*{Acknowledgements}
We wish to acknowledge financial support from an Early Career Award from the Department of Energy, Program Manager Dr.~Steven~Lee, award number DE-SC0020998.

\bibliographystyle{elsarticle-num-names}

\end{document}